\documentclass[twocolumn,amsmath,amssymb,prd]{revtex4-1}

\usepackage{graphicx}
\usepackage{pstricks}

\def\al{\alpha}
\def\be{\beta}
\def\ga{\gamma}
\def\de{\delta}
\def\ep{\epsilon}
\def\va{\varepsilon}
\def\ze{\zeta}
\def\et{\eta}
\def\th{\theta}

\def\ka{\kappa}
\def\la{\lambda}

\def\rh{\rho}

\def\si{\sigma}

\def\ta{\tau}
\def\up{\upsilon}
\def\ph{\phi}

\def\ch{\chi}
\def\ps{\psi}
\def\om{\omega}
\def\Ga{\Gamma}

\def\Th{\Theta}
\def\La{\Lambda}
\def\Si{\Sigma}

\def\Om{\Omega}

\def\da{^\dagger}
\def\tr{\textrm{tr}}
\def\pr{^\prime}
\def\Da#1{D_{\al_{#1}}}
\def\Dan{D_{\al_n}}
\def\Db#1{D_{\be_{#1}}}
\def\Dnn{D_{(n)}}
\def\Dn#1{D_{(n_{#1})}}
\def\Dnm{D_{(n_m)}}

\def\Qhat{\widehat Q}

\def\hc{\textrm{h.c.}}
\def\simn{\si_{\mu\nu}}
\def\scdots{\!\cdots\!}

\newcounter{tc1}\newcounter{tc2}
\newcounter{tr1}\newcounter{tr2}
\newlength{\h}
\def\newtableau#1#2{\psset{unit=12pt,linewidth=0.5pt}%
  \setlength{\h}{#2\psunit}\setlength{\h}{0.5\h}\addtolength{\h}{-0.3\psunit}
  \begin{pspicture}[shift=-\h](#1,#2)\small%
    \setcounter{tc1}{0}\setcounter{tc2}{1}%
    \setcounter{tr1}{#2}\setcounter{tr2}{#2}\addtocounter{tr1}{-1}%
    \psline(0,0)(0,#2)(#1,#2)}
\def\endtableau{\end{pspicture}}
\def\newbox#1{%
  \psline(\value{tc1},\value{tr1})(\value{tc2},\value{tr1})(\value{tc2},\value{tr2})%
  \rput(\value{tc1},\value{tr1}){\rput(0.5,0.5){#1}}
  \addtocounter{tc1}{1}\addtocounter{tc2}{1}}

\def\newrow{%
  \addtocounter{tr1}{-1}\addtocounter{tr2}{-1}%
  \setcounter{tc1}{0}\setcounter{tc2}{1}}

\def\co{{\cal O}}
\def\cg{{\cal G}}
\def\cL{{\cal L}}
\def\cl{{\cal L}}
\def\fl{F^{\textrm L}}
\def\Ftil{\widetilde{F}}

\def\half{{\textstyle{\frac 1 2}}}
\def\quar{{\textstyle{\frac 1 4}}}
\def\eigh{{\textstyle{\frac 1 8}}}

\def\lsim{\mathrel{\rlap{\lower4pt\hbox{\hskip1pt$\sim$}}
    \raise1pt\hbox{$<$}}}
\def\gsim{\mathrel{\rlap{\lower4pt\hbox{\hskip1pt$\sim$}}
    \raise1pt\hbox{$>$}}}

\def\prt{\partial}
\def\prtn{\prt_{(n)}}

\newcommand{\beq}{\begin{equation}}
\newcommand{\eeq}{\end{equation}}
\newcommand{\bea}{\begin{eqnarray}}
\newcommand{\eea}{\end{eqnarray}}
\newcommand{\rf}[1]{(\ref{#1})}
\newcommand{\nn}{\nonumber}

\def\etal{{\it et al.}}

\def\psb{\overline{\ps}{}}

\def\ov{\overline}

\def\PP{X}
\def\QQ{Y}
\def\ring#1{{\mathaccent'27 #1}}
\def\von{\ring{X}_1}
\def\vtw{\ring{X}_2}
\def\vth{\ring{Y}}
\def\Diso{\ring{D}}
\def\Dison{\Diso_{(n)}}
\def\piso{\ring{\prt}}
\def\pison{\piso_{(n)}}

\def\Im{\hbox{Im}\,}

\begin{document}

\title{Gauge field theories with Lorentz-violating operators 
of arbitrary dimension}

\author{
V.\ Alan Kosteleck\'y and Zonghao Li}

\affiliation{
Physics Department, Indiana University, 
Bloomington, IN 47405, USA}

\date{December 2018; 
published in Phys.\ Rev.\ D {\bf 99}, 056016 (2019)} 

\begin{abstract}
The classification of Lorentz- and CPT-violating operators
in nonabelian gauge field theories is performed.
We construct all gauge-invariant terms 
describing propagation and interaction in the action
for fermions and gauge fields.
Restrictions to the abelian, Lorentz-invariant, 
and isotropic limits are presented. 
We provide two illustrative applications of the results
to quantum electrodynamics and quantum chromodynamics.
First constraints on nonlinear Lorentz-violating effects in electrodynamics
are obtained using data from experiments on photon-photon scattering,
and corrections from nonminimal Lorentz and CPT violation 
to the cross section for deep inelastic scattering are derived.
\end{abstract}

\maketitle

\section{Introduction}

Nonabelian gauge theories,
introduced by Yang and Mills in 1954
\cite{ym54}, 
play a central role in physics.
Applications to particle physics
such as the Standard Model (SM)
typically combine nonabelian gauge invariance 
with the foundational Lorentz symmetry of relativity.  
In recent years,
however,
attention has been drawn to the possibility
that tiny violations of Lorentz symmetry
could arise in a unified theory of gravity and quantum physics
such as strings
\cite{ksp},
triggering many searches for potentially observable signals 
in laboratory experiments and astrophysical observations
\cite{tables}.
Studies of Lorentz-violating nonabelian gauge theories
are therefore of immediate interest
in the phenomenological context.

Using effective field theory
\cite{sw},
a realistic and comprehensive description of Lorentz violation
encompassing the nonabelian gauge symmetry of the SM
can be developed.
This approach starts with the SM action coupled to General Relativity (GR)
and adds all coordinate-independent terms formed as the contraction
of a Lorentz-violating operator with a coefficient
governing the size of its physical effects.
The resulting framework is 
called the Standard-Model Extension (SME)
\cite{ck,ak04}.
The SME also provides a general description 
of CPT violation in realistic field theory 
because CPT invariance follows from Lorentz invariance
in effective field theory
\cite{ck,owg}.

In Minkowski spacetime,
the SME is a Lorentz-violating nonabelian gauge theory
based on the gauge group SU(3)$\times$SU(2)$\times$U(1).
The subset of terms constructed from operators 
with mass dimensions $d\leq 4$ forms the minimal SME,
which is power-counting renormalizable.
In a scenario with the SM and GR 
emerging as the low-energy limit 
of a unified theory of quantum physics and gravity,
operators with smaller $d$
can be expected to dominate the low-energy Lorentz-violating physics.
The experimental viability of any specific Lorentz-violating model 
compatible with realistic effective field theory
can be determined by matching the model parameters 
to the corresponding subset of SME coefficients
and their known experimental bounds
\cite{tables,reviews}.

Most treatments of Lorentz violation to date 
emphasize effects in the minimal SME.
For the nonminimal sector,
a complete enumeration exists of operators at arbitrary $d$ 
that modify the gauge-invariant propagation of various particle species,
including scalars, Dirac fermions, neutrinos, photons, and gravitons 
\cite{km,ek18}.
However,
much less is known about the general form
of gauge-invariant interactions at arbitrary $d$.
In particular,
no general treatment of the nonabelian case exists to date, 
although a few works consider special
nonminimal nonabelian Lorentz-violating operators 
\cite{bp08,ar14,em14,jc15,em16,nvt16,mf17,mfm18}.
Even in the comparatively simple case 
of Lorentz-violating quantum electrodynamics (QED),
only a subset of interactions for $d\leq 6$ 
has been systematically classified
\cite{dk16}.

The focus of the present work is closing this gap in the literature.
We exhibit here a construction of Lorentz-violating terms at arbitrary $d$
in the Lagrange density for a generic nonabelian gauge field theory
with fermion couplings.
The case of Lorentz-violating QED emerges as the abelian limit
of this theory.
For definiteness,
we assume here a Minkowski background spacetime
and an action invariant both under a nonabelian gauge group
and under spacetime translations,
thereby insuring conservation of energy and momentum.
This setup generates a framework appropriate
for many experimental applications,
as we illustrate below with examples drawn from Lorentz-violating QED
and from the theory 
of Lorentz-violating quantum chromodynamics (QCD) and QED coupled to quarks. 
The operators constructed here 
are potentially of formal theoretical importance 
in contexts such as studies of causality and stability
\cite{causality}
and the recently uncovered links to Riemann-Finsler geometry
\cite{ek18,finsler}.
They also offer prospective applications 
to searches for Lorentz-invariant geometric forces extending GR
such as torsion
\cite{torsion}
and nonmetricity
\cite{nonmetricity},
as well as to studies of phenomenological Lorentz-violating scenarios 
focusing on operators with $d>4$
such as supersymmetric models 
\cite{susy}
and noncommutative quantum field theories 
\cite{ncqft}.

This work is organized as follows.
In Sec.\ \ref{Gauge-covariant operators},
we demonstrate that any gauge-covariant combination
of covariant derivatives and gauge-field strengths
can be expressed in a standard form,
and we derive some useful properties.
The general nonabelian gauge field theory
including Lorentz and CPT violation
is constructed in Sec.\ \ref{Gauge Field Theory}.
We present in turn the fermion sector,
the pure-gauge sector,
and the fermion-gauge sector of this theory,
providing in tabular form the explicit expressions
for all operators of mass dimension $d\leq 6$.
This section also considers several limits of interest,
including Lorentz-violating QED,
Lorentz-violating QCD and QED,
the Lorentz-invariant restriction,
and the isotropic case. 

Two experimental applications of the results
are considered in Sec.\ \ref{Experiments}.
One is the effect on light-by-light scattering
of certain nonminimal operators in Lorentz-violating QED.
Data from an experiment at the Large Hadron Collider (LHC)
are used to obtain first constraints 
on some coefficients for Lorentz violation.
The second application is to deep inelastic scattering (DIS).
Corrections to the cross section for electron-proton scattering
arising from certain nonminimal Lorentz- and CPT-violating operators 
are obtained.  
Throughout this work,
our conventions follow those of Ref.\ \cite{km}.
In particular,
the Minkowski metric $\et_{\mu\nu}$ has negative signature,
and the Levi-Civita tensor $\ep^{\ka\la\mu\nu}$ 
is defined with $\ep^{0123}=+1$.

\section{Gauge-covariant operators}
\label{Gauge-covariant operators}

One goal of this work is to classify and enumerate 
all terms in the Lagrange density
for a Lorentz- and CPT-violating nonabelian gauge theory.
For this purpose,
it is useful first to establish a standard basis 
spanning the space of all gauge-covariant operators.
A generic term in the Lagrange density
can then be decomposed in terms of the standard basis.
In this section,
we identify a suitable standard basis 
and establish some of its properties.

\subsection{Setup}
\label{Basics}

Consider a gauge field theory of a Dirac fermion $\ps$ 
taking values in a specific representation $U$
of the gauge group $\cg$.
Assuming $\cg$ is a compact Lie group,
a suitable inner product can be chosen 
to make the representation unitary 
\cite{ms07},
$U U\da = U\da U = 1$.
Under a gauge transformation,
$\ps$ transforms covariantly by construction,
$\ps \to U\ps$,
while its Dirac conjugate $\psb\equiv \ps\da \ga^0$ transforms as
$\psb \to \psb U\da$.

Acting on $\ps$,
the gauge-covariant derivative can be written as
$D_\mu\ps = \prt_\mu \ps-igA_\mu\ps$, 
where $g$ is the gauge coupling constant
and $A_\mu$ is the gauge field in the $U$ representation. 
Since $U$ is unitary, 
$A_\mu$ is hermitian.
By definition,
the covariant derivative satisfies 
$D_\mu \to U D_\mu U\da$
under a gauge transformation,
and so the gauge field obeys
$A_\mu \to U A_\mu U\da+(i/g)U \prt_\mu U\da$.

The commutator of two covariant derivatives acting on $\ps$
generates the gauge-field strength $F_{\mu\nu}$ in the $U$ representation,
${[D_\mu,D_\nu]} \ps = -ig F_{\mu\nu}\ps$.
Like $A_\mu$,
$F_{\mu\nu}$ is hermitian.
Under a gauge transformation,
$F_{\mu\nu} \to U F_{\mu\nu} U\da$.
The action of the covariant derivative on the field strength is
$D_\mu F_{\be\ga} = \prt_\mu F_{\be\ga} -ig [ A_\mu, F_{\be\ga}]$,
while the product of two covariant derivatives
acting on the field strength obeys 
$D_\mu(D_\nu F_{\be\ga}) - D_\nu(D_\mu F_{\be\ga}) 
= -ig [F_{\mu\nu}, F_{\be\ga}]$.

The above results imply that a gauge transformation
on single covariant derivatives of the fields yields
$D_\mu \ps \to U (D_\mu \ps)$ and
$\overline{D_\mu \ps} \to (\overline{D_\mu \ps}) U\da$
for fermions,
along with $D_\mu F_{\be\ga} \to U (D_\mu F_{\be\ga}) U\da$
for the gauge-field strength.
By induction,
we can prove that multiple covariant derivatives on each field
transform according to
\bea
\Da1\scdots\Dan\ps &\to& U(\Da1\scdots\Dan\ps),
\nn\\
\overline{\Da1\scdots\Dan\ps} &\to& (\overline{\Da1\scdots\Dan\ps})U\da,
\nn\\
\Da1\scdots\Dan F_{\be\ga} &\to& U(\Da1\scdots\Dan F_{\be\ga})U\da.
\eea

It is convenient to call an operator $\co$ a gauge-covariant operator
if $\co \to U \co U\da$ under a gauge transformation.
Gauge-covariant operators are of interest in the present context
because they can be used to construct gauge-invariant terms
such as $\tr(\co)$ and $(\ov{\co_1 \ps}) \co_2 \ps$,
which are therefore candidates for inclusion in the general Lagrange density
for the Lorentz-violating gauge field theory.
In particular,
we see that any operator $\co$ formed as a mixture of $D$ and $F$
is gauge covariant.

\subsection{Result}
\label{proof}

In principle,
one could construct the desired general Lagrange density
for a Lorentz- and CPT-violating gauge field theory
by adding all possible gauge-invariant operators
to the usual minimal-coupling terms.
However,
this procedure would introduce substantial redundancies
due to relations among the gauge-invariant operators. 
Instead,
we can characterize gauge-covariant operators in terms of a standard basis set,
which can then be used to construct the Lagrange density 
with controlled or no redundancy.

The key result is that the gauge-covariant operators 
formed as any mixture of $D$ and $F$ 
can be expressed as a linear combination of operators 
of the standard form 
\beq
(D_{(n_1)} F_{\mu_1 \nu_1}) (D_{(n_2)} F_{\mu_2 \nu_2})
\cdots (D_{(n_m)} F_{\mu_m \nu_m}) D_{(n_{m+1})},
\label{form}
\eeq
where $D_{(n)} = (1/n!) \sum \Da1 \Da2 \scdots D_{\al_n}$
with the summation performed over all permutations 
of $\al_1, \al_2, \cdots \al_n$.
In what follows,
we prove this result and present some of its properties.

\subsubsection{Proof}

For the proof,
we suppose that the operator \rf{form} acts on $\ps$.
An analogous argument applies for the case 
where the operator acts on $F$ instead.

Given a gauge-covariant operator $\co$ formed
as an arbitrary mixture of $D$ and $F$,
we first use standard product rules like
$D_\mu F_{\al\be} D_\nu = (D_\mu F_{\al\be}) D_\nu + F_{\al\be} D_\mu D_\nu$
to express it as a linear combination
of terms in the block form
\beq
\label{eq:block}
\co = \sum (D D \cdots D F) \cdots (D D \cdots D F) D D \cdots D,
\eeq
where all subscripts are omitted for simplicity.
Each covariant derivative $D$ then acts on at most one field strength $F$.
To prove the main result,
it therefore suffices to show 
that terms like $D_{\al_1} \Da2 \cdots D_{\al_n}$
can be expressed as linear combinations of the form \rf{form}.

We use mathematical induction.
For $n=1$, 
$\Da1$ already takes the form \rf{form}.
Assume that for $n \leq k$ 
the term $D_{\al_1} \Da2 \cdots D_{\al_n}$
can be written as a linear combination of terms having the form \rf{form}.
Then,
for $n=k+1$ we must consider 
$D_{\al_1} D_{\al_2} \cdots D_{\al_k} D_{\al_{k+1}}$.
Replacing the first $k$ derivatives
by the appropriate linear combination 
of terms having the form \rf{form}
yields expressions with the structure 
$(D_{(n_1)} F_{\be_1\ga_1}) \cdots D_{(n_{m+1})} D_{\al_{k+1}}$.
This shows that it suffices to focus only on the term 
$D_{(n_{m+1})} D_{\al_{k+1}}$.

If $n_{m+1}\leq k-1$, 
then we get at most $k$ covariant derivatives.
By induction, 
all such terms can be written as a linear combination 
of terms having the form \rf{form}.

For $n_{m+1}=k$, 
we must consider $D_{(k)} D_{\al_{k+1}}$.
The index symmetries of this product can be displayed
using Young tableaux,
\beq
\newtableau{4}{1}
\newbox{$\scriptstyle 1$} \newbox{$\scriptstyle 2$} \newbox{$\scriptstyle \cdots$} \newbox{$\scriptstyle k$}
\endtableau
\otimes
\newtableau{1}{1}
\newbox{$\scriptstyle k\!+\!1$}
\endtableau
=
\newtableau{5}{1}
\newbox{$\scriptstyle 1$} \newbox{$\scriptstyle 2$} \newbox{$\scriptstyle \cdots$} \newbox{$\scriptstyle k$}\newbox{$\scriptstyle k\!+\!1$}
\endtableau
\oplus
\newtableau{4}{2}
\newbox{$\scriptstyle 1$} \newbox{$\scriptstyle 2$} \newbox{$\scriptstyle \cdots$} \newbox{$\scriptstyle k$}
\newrow \newbox{$\scriptstyle k\!+\!1$}
\endtableau .
\eeq
The first Young tableau on the right-hand side represents $D_{(k+1)}$, 
which has the claimed form. 
We can therefore limit further consideration to the second Young tableau.

We can prove that the expression involving the second Young tableau 
contains least one factor of the field strength $F$.
Explicitly,
we can write this expression in the form
\bea
&& 
\hskip-30pt
D_{(\al_1}D_{\al_2}\cdots D_{\al_k)}D_{\al_{k+1}}
-D_{(\al_{k+1}}D_{\al_2}\cdots D_{\al_k)}D_{\al_1}
\nn\\
&&
\hskip 20pt
= \frac2k \sum_{t=0}^{k-1} 
D_{(t)}D_{[\al_1}D_{(k-t-1)}D_{\al_{k+1}]},
\label{young}
\eea
where the parentheses around indices indicate symmetrization
with a factor of $1/k!$
and the brackets indicate antisymmetrization 
on $\al_1$ and $\al_{k+1}$ with a factor of $1/2$.
Note that this result is symmetric 
under the interchange of the indices in $D_{(t)}$ 
with those in $D_{(k-t-1)}$.

A specific term inside the sum in Eq.\ \rf{young} takes the form
\bea
D_{(t)}D_{[\al_1}D_{(k-t-1)}D_{\al_{k+1}]}
&=&
D_{(t)}D_{[\al_1}D_{\al_{k+1}]}D_{(k-t-1)}
\nn\\
&&
+D_{(t)}D_{[\al_1}[D_{(k-t-1)},D_{\al_{k+1}]}].
\nn\\
\eea
The first term on the right-hand side
involves $[D_{\al_1},D_{\al_{k+1}}]=-igF_{\al_1 \al_{k+1}}$,
thus confirming the appearance of a factor of $F$.
For the second term,
we note that
\bea
&&
\hskip-20pt
D_{(t)}D_{\al_1}[D_{(k-t-1)},D_{\al_{k+1}}]
\nn\\
&=&
\tfrac 1 {(k-t-1)!} \sum 
D_{(t)}D_{\al_1}[\Db1 \Db2 \cdots D_{\be_{k-t-1}},D_{\al_{k+1}}]
\nn\\
&=&
\tfrac 1 {(k-t-1)!} \sum 
D_{(t)}D_{\al_1}\big(\Db1 [\Db2 \cdots D_{\be_{k-t-1}},D_{\al_{k+1}}]
\nn\\
&&
\hskip 50pt
+[\Db1 ,D_{\al_{k+1}}] \Db2 \cdots D_{\be_{k-t-1}}\big)
\nn\\
&=&
\tfrac 1 {(k-t-1)!} \sum 
\big(D_{(t)}D_{\al_1} \Db1 [\Db2 \cdots D_{\be_{k-t-1}},D_{\al_{k+1}}]
\nn\\
&&
\hskip 50pt
-igD_{(t)}D_{\al_1}F_{\be_1\al_{k+1}} \Db2 \cdots D_{\be_{k-t-1}}\big),
\nn\\
\eea
where the summations are over all permutations 
of $\be_1, \be_2, \cdots \be_{k-t-1}$.
Continuing this reductive process yields a string of terms,
each of which contains one factor of $F_{\mu\nu}$ 
along with $(k-1)$ covariant derivatives. 
Therefore, 
every term in the expression 
for the second Young tableau in Eq.\ \rf{young}
involves at least one field strength $F$.
Each such term has the general form 
$D_{\ga_1} \cdots D_{\ga_s}F_{\mu\nu}D_{\ga_{s+1}} \cdots D_{\ga_{k-1}}$.
Using product rules as before, 
we can transform this into the block form \rf{eq:block}.
In each block,
we have at most $k-1 \leq k$ covariant derivatives.
By the induction assumption,
these terms can all be written as linear combinations 
having the form \rf{form}.
This completes the induction step and proves the claimed result.

\subsubsection{Linear independence}

The above argument shows that the basis \rf{form}
of gauge-covariant operators is complete,
but it leaves open the question of linear independence.
The situation in this respect can differ
depending on whether the group $\cg$ is abelian or nonabelian.

Consider first an abelian gauge field theory,
for which
$D_{(n)} F_{\be\ga} = \prt_{(n)} F_{\be\ga}$.
Note that
$\prt_{(n_1)} F_{\be_1 \ga_1}$
commutes with 
$\prt_{(n_2)} F_{\be_2 \ga_2}$
and also that the operators $\{ \prt_{(n)} F_{\be\ga} \}$ 
are linearly independent for different $n$.
A complete set of linearly independent basis operators 
can therefore be selected as 
\beq
\{ (\prt_{(n_1)} F_{\be_1 \ga_1}) \cdots 
(\prt_{(n_t)} F_{\be_t \ga_t}) D_{(n_{t+1})} 
| n_1 \leq \cdots \leq n_t \}.
\label{abelian}
\eeq
This set can be used to classify all gauge-covariant operators 
formed from mixtures of $D$ and $F$ in QED,
in a manner free of redundancy.

For a nonabelian group,
the noncommutativity of products of the field strength $F$
implies that gauge-covariant operators having the form \rf{form} 
are generically linearly independent.
When $\cg$ is SU(2),
for example,
the three generators of $\cg$ are mutually noncommutative,
implying that basis elements of the form \rf{form} are linearly independent.

However,
if a nonabelian $\cg$ has an abelian Lie subgroup of dimension two or more,
the choice of linearly independent basis becomes more intricate.
In QCD,
for example,
$U$ is the vector representation
of the gauge group SU(3),
which has a subgroup U(1)$\times$U(1) homomorphic to the torus $T^2$. 
This subgroup is spanned by 3$\times$3 commuting matrices 
with diagonal elements of the form 
$\{ e^{i \th_1}, e^{i \th_2}, e^{-i (\th_1+\th_2)} \}$.
Gauge-covariant operators for basis elements in U(1)$\times$U(1) 
can therefore be taken to have the form \rf{abelian},
while other basis elements are of the form \rf{form}.
This shows that in a nonabelian gauge field theory
the selection of a basis for the gauge-covariant operators
depends on the gauge group.
For simplicity in what follows,
we adopt basis elements of the form \rf{form} 
for nonabelian gauge field theory
and tolerate any ensuing partial redundancy.

\subsubsection{Hermitian conjugation}
\label{sec:hermitian}

For some of the applications to follow,
it is useful to have explicit expressions
for the hermitian conjugates 
of the component terms in the gauge-covariant operators \rf{form}
and of general fermion bilinears.
We obtain these next.

First,
we prove the identity 
\beq
(\Dnn F_{\be\ga})\da=\Dnn F_{\be\ga}.
\label{conj}
\eeq
To begin,
use direct calculation to show
$(D_\al F_{\be\ga})\da = D_\al F_{\be\ga}$.
We can then use mathematical induction
to show that
$(\Da1\scdots\Dan F_{\be\ga})\da=\Da1\scdots\Dan F_{\be\ga}$
for any $n$,
as follows.
The result is true for $n=1$.
Assuming it holds for $n=k$,
we find 
\bea
&& 
\hskip-30pt
(D_\al \Da1\scdots D_{\al_k} F_{\be\ga})\da
\nn\\
&=&
\prt_\al \Da1\scdots D_{\al_k} F_{\be\ga}
+ig[\Da1\scdots D_{\al_k} F_{\be\ga},A_\al]
\nn\\
&=& 
D_\al \Da1\scdots D_{\al_k} F_{\be\ga}.
\eea
The result therefore also holds for $n=k+1$, as required.
The desired identity \rf{conj} 
then follows by symmetrization on indices.

For conjugation of expressions involving fermions,
two results are useful.
For one,
the identity \rf{conj} directly implies
\beq
\ov{(D_{(n)}F_{\be\ga})\ps}=\psb (D_{(n)}F_{\be\ga}) .
\eeq
The other identity involves hermitian conjugation
of a general gauge-invariant bilinear of the form
$(\ov{\co_1 \ps})\Ga_I(\co_2 \ps)$,
where $\co_1$ and $\co_2$ are gauge-covariant operators 
and the spinor space is spanned by 
the 16 matrices 
$\Ga_I \in \{ 1, i \ga_5, \ga_\mu, \ga_5 \ga_\mu, \simn/2 \}$.
Following the argument in Sec.\ \ref{proof},
we can choose $\co_1$ and $\co_2$ 
to have the form \rf{form}
and thereby write the general bilinear as 
\beq
(\ov{\Dn0 \ps}) \Ga_I (\Dn1 F_{\be_1 \ga_1})\cdots 
(\Dnm F_{\be_m \ga_m})(D_{(n_{m+1})}\ps).
\label{bilinear}
\eeq

To find the hermitian conjugate of the bilinear \rf{bilinear},
note that
$(D_\mu\ps)\da=(\ov{D_\mu\ps})\ga_0$ 
and
$(\ov{D_\mu\ps})\da=\ga_0 (D_\mu \ps)$.
In analogy with the proof of the identity \rf{conj},
we can use mathematical induction to show the two results 
$(\Dnn\ps)\da=(\ov{\Dnn\ps}) \ga_0$
and 
$(\ov{\Dnn\ps})\da=\ga_0 (\Dnn\ps)$.
Recalling the relation $\Ga_I=\ga_0 \Ga_I\da \ga_0$,
we finally arrive at the elegant conjugation identity 
\bea
&&
\hskip -12pt
[ (\ov{\Dn0 \ps}) \Ga_I (\Dn1 F_{\be_1 \ga_1})
\scdots (\Dnm F_{\be_m \ga_m})(D_{(n_{m+1})}\ps) ]\da 
\nn\\
&=&
(\ov{D_{(n_{m+1})} \ps}) \Ga_I (\Dnm F_{\be_m \ga_m})
\scdots (\Dn1 F_{\be_1 \ga_1})(\Dn0\ps).
\nn\\
\eea

\section{Gauge Field Theory}
\label{Gauge Field Theory}

To construct the general gauge-invariant theory 
including Lorentz and CPT violation,
it is convenient to separate the full Lagrange density $\cL$
into three parts,
\beq
\cL = \cL_\ps + \cL_A + \cl\pr,
\label{fullL}
\eeq
where $\cl_\ps$ denotes the fermion sector 
including covariant-derivative couplings,
$\cL_A$ denotes the pure-gauge sector,
and $\cl\pr$ is a fermion-gauge sector
containing products of operators appearing in the first two parts. 
In this section,
each of these pieces of $\cL$ are considered in turn.
We offer some observations about the general situation 
and provide explicit expressions for terms of low mass dimensionality.

\subsection{Fermion sector}
\label{spinor}

For a Dirac fermion,
the general gauge-invariant Lagrange density $\cl_\ps$ 
involving self interactions and gauge interactions via covariant derivatives
can be expanded in a series of the schematic form
\bea
\cL_\ps&=&\half \psb (D_\mu \ga^\mu -m)\ps +\hc
\nn\\
&&
+\psb \Qhat \ps +\hc
\nn\\
&&
+ (\psb \Qhat_1 \ps) \big( (\ov{D_{(m_1)}\ps}) \Qhat_2 \ps \big)+\hc
\nn\\
&&
+(\psb \Qhat_3 \ps)  \big( (\ov{D_{(m_2)}\ps}) \Qhat_4 \ps \big)
\big( (\ov{D_{(m_3)}\ps}) \Qhat_5 \ps \big)+\hc
\nn\\
&&
+\cdots,
\label{fermion}
\eea
where $\Qhat$, $\Qhat_1$, $\Qhat_2$, $\cdots$ 
are understood to be gauge-covariant operators.
Adopting the results in Sec.\ \ref{Gauge-covariant operators},
each of these operators can be expressed as a linear combination
of operators of the form
\beq
\Ga_I (D_{(n_1)} F_{\be_1\ga_1}) 
\cdots (D_{(n_m)} F_{\be_m\ga_m})  D_{(n_{m+1})}.
\label{genop}
\eeq
For brevity,
the expression \rf{fermion} for $\cl_\ps$ omits spacetime indices 
in all but the conventional Dirac term
and also omits all coefficients for Lorentz violation.
It is understood that each term in $\cL_\ps$
comes with a coefficient having spacetime indices
contracted with all free spacetime indices 
on the total operator for that term,
thereby maintaining observer Lorentz symmetry of the theory
\cite{ck}.
Note also that no derivative $\Dnn$ acts on $\psb$ 
in the first bilinear of each unconventional term in $\cl_\ps$.
Any terms with these derivatives are equivalent
to the displayed ones via partial integration,
modulo possible surface terms in the action.

\renewcommand\arraystretch{1.4}
\begin{table*}
\caption{
\label{tab:spinor}
Terms of low mass dimension $d\leq 6$ 
in the fermion Lagrange density $\cl_\ps$.}
\setlength{\tabcolsep}{8pt}
\begin{tabular}{cl}
\hline
\hline
Component & Expression \\
\hline
$	\cL_{\ps0}	$	&		$		\half \psb (\ga^\mu i D_\mu - m_\ps ) \ps	+	\hc	$	\\	[2pt]
$	\cL^{(3)}_\ps	$	&		$	-	i m_5 \psb \ga_5\ps -a^\mu\psb\ga_\mu\ps					
						-	b^\mu\psb\gamma_5\ga_\mu\ps					
						-	\half H^{\mu\nu}\psb\simn \ps			$	\\	[2pt]
$	\cL^{(4)}_\ps	$	&		$		\half c^{\mu\al}\psb\ga_\mu i D_\al \ps					
						+	\half d^{\mu\al}\psb \ga_5 \ga_\mu i D_\al \ps					
						+	\half e^\al \psb iD_\al \ps					
						+	\half i f^\al \psb \ga_5 iD_\al \ps					
						+	\quar g^{\mu\nu\al} \psb \simn iD_\al \ps	+	\hc	$	\\	[2pt]
$	\cL_{\ps D}^{(5)}	$	&		$	-	\half m^{(5)\al\be}\psb iD_{(\al} iD_{\be)}\ps					
						-	\half i m_5^{(5)\al\be} \psb \ga_5 iD_{(\al} iD_{\be)} \ps			$	\\	[2pt]
			&	\hskip20pt	$	-	\half a^{(5)\mu\al\be} \psb \ga_\mu iD_{(\al} iD_{\be)} \ps					
						-	\half b^{(5)\mu\al\be}\psb \ga_5 \ga_\mu iD_{(\al} iD_{\be)} \ps					
						-	\quar H^{(5)\mu\nu\al\be} \psb \simn iD_{(\al} iD_{\be)} \ps	+	\hc	$	\\	[2pt]
$	\cL_{\ps F}^{(5)}	$	&		$	-	\half m_F^{(5)\al\be} \psb F_{\al\be} \ps					
						-	\half im_{5F}^{(5)\al\be} \psb \ga_5 F_{\al\be} \ps					
						-	\half a_F^{(5)\mu\al\be} \psb \ga_\mu F_{\al\be} \ps					
						-	\half b_F^{(5)\mu\al\be} \psb \ga_5 \ga_\mu F_{\al\be} \ps					
						-	\quar H_F^{(5)\mu\nu\al\be} \psb \simn F_{\al\be} \ps			$	\\	[2pt]
$	\cL_{\ps D}^{(6)}	$	&		$		\half c^{(6)\mu\al\be\ga} \psb \ga_\mu iD_{(\al} iD_\be iD_{\ga)} \ps					
						+	\half d^{(6)\mu\al\be\ga} \psb \ga_5 \ga_\mu iD_{(\al} iD_\be iD_{\ga)} \ps			$	\\	[2pt]
			&	\hskip20pt	$	+	\half e^{(6)\al\be\ga} \psb iD_{(\al} iD_\be iD_{\ga)} \ps					
						+	\half i f^{(6)\al\be\ga} \psb \ga_5 iD_{(\al} iD_\be iD_{\ga)} \ps					
						+	\quar g^{(6)\mu\nu\al\be\ga} \psb \simn iD_{(\al} iD_\be iD_{\ga)} \ps	+	\hc	$	\\	[2pt]
$	\cL_{\ps F}^{(6)}	$	&		$		\quar c_F^{(6)\mu\al\be\ga} \psb \ga_\mu F_{\be\ga} iD_\al \ps					
						+	\quar d_F^{(6)\mu\al\be\ga} \psb \ga_5 \ga_\mu F_{\be\ga} iD_\al \ps			$	\\	[2pt]
			&	\hskip20pt	$	+	\quar e_F^{(6)\al\be\ga} \psb F_{\be\ga} iD_\al \ps					
						+	\quar i f_F^{(6)\al\be\ga} \psb \ga_5 F_{\be\ga} iD_\al \ps					
						+	\eigh g_F^{(6)\mu\nu\al\be\ga} \psb \simn F_{\be\ga} iD_\al \ps	+	\hc	$	\\	[2pt]
$	\cL_{\ps DF}^{(6)}	$	&		$	-	\half m_{DF}^{(6)\al\be\ga} \psb (D_\al F_{\be\ga})\ps					
						-	\half i m_{5DF}^{(6)\al\be\ga} \psb \ga_5 (D_\al F_{\be\ga})\ps			$	\\	[2pt]
			&	\hskip20pt	$	-	\half a_{DF}^{(6)\mu\al\be\ga} \psb \ga_\mu (D_\al F_{\be\ga})\ps					
						-	\half b_{DF}^{(6)\mu\al\be\ga} \psb \ga_5 \ga_\mu (D_\al F_{\be\ga})\ps					
						-	\quar H_{DF}^{(6)\mu\nu\al\be\ga} \psb \simn (D_\al F_{\be\ga})\ps			$	\\	[2pt]
$	\cL_{\ps\ps}^{(6)}	$	&		$		k_{SS}(\psb\ps)(\psb\ps)					
						-	k_{PP}(\psb\ga_5\ps)(\psb\ga_5\ps)					
						+	ik_{SP}(\psb\ps)(\psb\ga_5\ps)			$	\\	[2pt]
			&	\hskip20pt	$	+	(k_{SV})^\mu(\psb\ps)(\psb\ga_\mu\ps)					
						+	(k_{SA})^\mu (\psb\ps) (\psb\ga_5\ga_\mu\ps)					
						+	\half (k_{ST})^{\mu\nu}(\psb\ps)(\psb\simn\ps)			$	\\	[2pt]
			&	\hskip20pt	$	+	i(k_{PV})^\mu (\psb\ga_5\ps) (\psb\ga_\mu\ps)					
						+	i(k_{PA})^\mu (\psb\ga_5\ps) (\psb\ga_5\ga_\mu\ps)					
						+	\half i(k_{PT})^{\mu\nu} (\psb\ga_5\ps) (\psb\simn\ps)			$	\\	[2pt]
			&	\hskip20pt	$	+	\half (k_{VV})^{\mu \nu}(\psb\ga_\mu\ps)(\psb\ga_\nu\ps)					
						+	\half (k_{AA})^{\mu \nu}(\psb\ga_5\ga_\mu\ps)(\psb\ga_5\ga_\nu\ps)					
						+	(k_{V\!A})^{\mu\nu}(\psb\ga_\mu\ps) (\psb \ga_5\ga_\mu \ps)			$	\\	[2pt]
			&	\hskip20pt	$	+	\half (k_{VT})^{\la\mu\nu} (\psb \ga_\la \ps) (\psb \simn \ps)					
						+	\half (k_{AT})^{\la\mu\nu} (\psb \ga_5\ga_\la \ps) (\psb \simn \ps)					
						+	\eigh (k_{TT})^{\ka\la\mu\nu}(\psb\si_{\ka\la}\ps)(\psb\simn\ps)			$	\\	[2pt]
\hline
\hline
\end{tabular}
\end{table*}

For practical applications and to obtain explicit expressions,
it is useful to expand $\cl_\ps$
in a series organized according to the mass dimensions of the operators.
This series can be written as
\beq
\cL_\ps=
\cL_{\ps0} +\cL^{(3)}_\ps+\cL^{(4)}_\ps+\cL^{(5)}_\ps+\cL^{(6)}_\ps+\cdots,
\eeq
where $\cL_{\ps0}$ is the conventional Dirac Lagrange density
and where the superscript $d$ on $\cl^{(d)}$ denotes the mass dimension
of operators included in $\cl^{(d)}$.

Table \ref{tab:spinor} displays
all terms appearing in the fermion Lagrange density $\cl_\ps$
with mass dimension six or less.
The first column of the table lists the components of the Lagrange density.
For $d=3$ and 4,
the terms displayed are power-counting renormalizable
and match the corresponding ones found 
in the nonabelian sector of the minimal SME
\cite{ck}.
The component of the Lagrange density with $d=5$ is split into two pieces,
$\cL^{(5)}_\ps = \cL_{\ps D}^{(5)} + \cL_{\ps F}^{(5)}$.
The first contains terms involving only symmetrized covariant derivatives,
while the second contains ones involving the gauge-field strength.
Analogously,
it is convenient to separate the component of $\cl_\ps$ with $d=6$
into four pieces, 
$\cL^{(6)}_\ps = \cL_{\ps D}^{(6)} +\cL_{\ps F}^{(6)}
+\cL_{\ps DF}^{(6)} +\cL_{\ps\ps}^{(6)}$,
by grouping terms with related structures.
Except for $\cL_{\ps\ps}^{(6)}$,
the terms with $d=5$ and 6 are nonabelian generalizations
of ones already characterized in the QED context
\cite{dk16}.

The second column of the table presents the explicit expressions
for the terms in the Lagrange density.
To match conventional notation used in the literature in the QED limit 
\cite{km},
the coefficients in $\Qhat$
associated with the spinor matrices
$1$, $i \ga_5$, $\ga_\mu$, $\ga_5 \ga_\mu$, and $\simn/2$ 
in operators of odd mass dimension
are denoted by $-m$, $-m_5$, $-a^\mu$, $-b^\mu$, and $-H^{\mu\nu}$,
respectively,
while those in operators of even mass dimension
are denoted by $e$, $f$, $c^\mu$, $d^\mu$, and $g^{\mu\nu}$.
Following standard usage,
the dimension $d$ is omitted on coefficients for $d=3$ and 4. 
The labeling of spacetime indices is chosen 
so that the indices  
$\al$, $\be$, $\ldots$ correspond to gauge couplings
while $\mu$, $\nu$ correspond to spin properties. 
Parentheses enclosing $n$ spacetime indices 
represent total symmetrization with a factor of $1/n!$.
Other symmetries can be determined by inspection.
The abbreviation $\hc$ appearing at the end of some entries in the table 
implies the addition of the hermitian conjugate 
of all previous terms in the expression.

Each displayed term in the Lagrange density 
is the contraction of a coefficient 
and an operator invariant under nonabelian gauge transformations.
Both Lorentz-invariant and Lorentz-violating terms are included.
Physical effects violating CPT
are produced by terms containing operators 
with an odd number of spacetime indices,
all of which violate Lorentz invariance
\cite{ck,owg}.
All coefficients can be taken as real constants 
in an inertial frame in the vicinity of the Earth
\cite{ck,ak04},
and each coefficient has dimension GeV$^{4-d}$.
Note that the conventional mass term is included
as part of $\cl_{\ps 0}$ rather than part of $\cL^{(3)}_\ps$.
Also,
the component of the coefficient $c^{\mu\al}$ 
proportional to $\et^{\mu\al}$ 
represents a Lorentz-invariant contribution
that acts merely to renormalize 
the conventional Lorentz-invariant kinetic term in $\cl_{\ps 0}$,
so $c^{\mu\al}$ can be chosen traceless
without loss of generality.

\renewcommand\arraystretch{1.4}
\begin{table*}
\caption{
\label{tab:pure-gauge}
Terms of low mass dimension $d\leq 8$ 
in the pure-gauge Lagrange density $\cl_A$.}
\setlength{\tabcolsep}{8pt}
\begin{tabular}{cl}
\hline
\hline
Component & Expression \\
\hline
$	\cL_{A0}	$	&		$	\propto -\tr(F_{\mu\nu}F^{\mu\nu})	$	\\	[2pt]
$	\cL_A^{(1)}	$	&		$		k^{(1)\mu} \tr(A_\mu)	$	\\	[2pt]
$	\cL_A^{(3)}	$	&		$		k^{(3)\ka} \ep_{\ka\la\mu\nu} \tr (A^\la F^{\mu\nu} +\tfrac{2}{3} ig A^\la A^\mu A^\nu)	$	\\	[2pt]
$	\cL_A^{(4)}	$	&		$	-	\half k^{(4)\ka\la\mu\nu} \tr (F_{\ka\la} F_{\mu\nu})	$	\\	[2pt]
$	\cL_A^{(5)}	$	&		$	-	\half k^{(5)\al\ka\la\mu\nu} \tr (F_{\ka\la} D_\al F_{\mu\nu})	$	\\	[2pt]
$	\cL_A^{(6)}	$	&		$	-	\half k_D^{(6)\al\be\ka\la\mu\nu} \tr (F_{\ka\la} D_{(\al} D_{\be)} F_{\mu\nu})			
						-	\tfrac 1 6 k_F^{(6)\ka\la\mu\nu\rh\si} \tr( F_{\ka\la} F_{\mu\nu} F_{\rh\si})	$	\\	[2pt]
$	\cL_A^{(7)}	$	&		$	-	\half k_D^{(7)\al\be\ga\ka\la\mu\nu} \tr (F_{\ka\la} D_{(\al} D_\be D_{\ga)} F_{\mu\nu})			
						-	\tfrac 1 6 k_F^{(7)\al\ka\la\mu\nu\rh\si} \tr (F_{\ka\la} F_{\mu\nu} D_\al F_{\rh\si})	$	\\	[2pt]
$	\cL_A^{(8)}	$	&		$	-	\half k_D^{(8)\al\be\ga\de\ka\la\mu\nu} \tr( F_{\ka\la} D_{(\al} D_\be D_\ga D_{\de)} F_{\mu\nu})			
						-	\tfrac 1 6 k_{DF}^{(8)\al\be\ka\la\mu\nu\rh\si} \tr \big( F_{\ka\la}(D_\al F_{\mu\nu})(D_\be F_{\rh\si}) \big)	$	\\	[2pt]
			&	\hskip20pt	$	-	\tfrac 1 {24} k_F^{(8) \ka\la\mu\nu\rh\si\ta\up} \tr(F_{\ka\la}F_{\mu\nu}F_{\rh\si}F_{\ta\up})			
						-	\tfrac 1 {24} k_{FF}^{(8) \ka\la\mu\nu\rh\si\ta\up} \tr(F_{\ka\la}F_{\mu\nu})\tr(F_{\rh\si}F_{\ta\up})	$	\\	[2pt]
\hline
\hline
\end{tabular}
\end{table*}

We remark in passing that the freedom to adopt 
suitable canonical variables by judicious use of field redefinitions
implies that some of the terms shown in the table
may describe the same effects as others
and hence can be removed without changing the physics.
A simple example is the term $-i m_5 \psb \ga_5\ps$,
which modulo possible anomalies
can be absorbed into the conventional mass term
by a chiral redefinition of the Dirac field.
A general study of the implications of field redefinitions 
would be of interest but lies outside our present focus.

\subsection{Pure-gauge sector}
\label{pure-gauge}

In the pure-gauge sector,
gauge-invariant terms in the Lagrange density $\cl_A$
can be constructed as traces $\tr(\co)$ in the group space 
\cite{cp76}
of gauge-covariant operators $\co$
containing mixtures of covariant derivatives and field strengths.
Assuming the coefficients in the Lagrange density $\cl_A$ are constants
and recalling the form \rf{form} for arbitrary gauge-covariant operators,
we find that the structure of a generic gauge-invariant term 
for the pure-gauge sector can be written schematically as
\beq
\cl_A \supset
k^{\cdots} \big(\tr \big[ (DF) \scdots (DF) \big] \big)
\scdots \big(\tr \big[ (DF) \scdots (DF) \big] \big),
\label{puregauge}
\eeq
where all spacetime indices are suppressed
and $D$ denotes $\Dnn$.
Note that the cyclic property of the trace
implies that the factor $D_{(n_{m+1})}$
in the form \rf{form} is irrelevant here.
Note also that some terms \rf{puregauge}
may be related to others up to a surface term 
via the identity
$\tr(\co_1 D_\mu \co_2) = - \tr [ (D_\mu \co_1) \co_2 ]$,
which holds for any two gauge-covariant operators $\co_1$ and $\co_2$.

In the expression \rf{puregauge},
$k^{\cdots}$ denotes a generic coefficient
having indices understood to be contracted
with all indices on the factors of $D$ and $F$,
thus insuring observer Lorentz invariance of $\cl_A$.
For some terms,
hermiticity imposes a symmetry on the indices.
The traces are understood to be taken
in the $U$ representation of the gauge group $\cg$.
If $U$ is a reducible representation of a semisimple Lie group,
then the trace can be any combination of traces 
taken in the irreducible subrepresentations.

In addition to gauge-invariant operators,
the Lagrange density $\cl_A$ can also contain operators 
that generate surface terms under a gauge transformation.
Although these operators cause 
the Lagrange density to violate gauge invariance,
they nonetheless leave invariant the action
and are therefore also of interest in the present context.
Some of these operators fall outside the construction \rf{puregauge}
and must thus be obtained separately,
as described below.

The Lagrange density $\cl_A$ can be separated
into components containing operators of fixed mass dimension $d$.
It is convenient to write $\cl_A$ in the form
\beq
\cl_A=
\cL_{A0}+\cL^{(1)}_A+\cL^{(2)}_A+\cL^{(3)}_A+\cL^{(4)}_A
+\cL^{(5)}_A+\cdots,
\label{pglag}
\eeq
where $\cl_{A0}$ is the conventional Yang-Mills Lagrange density
in the $U$ representation
and the superscripts denote the operator mass dimension.

Table \ref{tab:pure-gauge} 
displays terms in the pure-gauge Lagrange density 
with mass dimension $d\leq 8$.
The components of the Lagrange density $\cl_A$
are listed in the first column,
while the second column contains the explicit expressions
formed via contractions of coefficients and field operators.
Note that both Lorentz-invariant and Lorentz-violating terms appear,
with the former emerging by specifying coefficients 
purely in terms of the Minkowski metric and Levi-Civita tensor.
Any term involving an operator 
with an odd number of spacetime indices
is odd under CPT.

The coefficients appearing in the table have dimension GeV$^{4-d}$
and can be assumed real and constant in an inertial frame near the Earth 
\cite{ck,ak04}.
The notation for coefficients adopted in the table 
is generic and can be used for any gauge group $\cg$.
Standard alternative notations used in the literature 
for coefficients with $d\leq 4$
appear in physical applications involving certain limiting cases.
For each term in the second column of the table,
the spacetime indices are chosen so that
$\al$, $\be$, $\ldots$ 
are associated with symmetrized covariant derivatives  
and $\ka$, $\la$, $\ldots$ with field strengths. 
Parentheses around a set of $n$ spacetime indices
imply total symmetrization with a factor of $1/n!$.
Other index symmetries are inherited
from the antisymmetry of $F_{\mu\nu}$
or the cyclic property of the trace.
For example,
the identity
$\tr( F_{\ka\la} \Dnn F_{\mu\nu}) = - \tr( F_{\mu\nu} \Dnn F_{\ka\la})$
implies a corresponding antisymmetry property 
for odd-$d$ terms quadratic in the field strength. 

Some of the contributions at mass dimensions $d\leq 3$
lie outside the construction \rf{puregauge}.
The component at mass dimension one involves the gauge field $A_\mu$ 
and is gauge invariant up to a surface term 
provided $k^{(1)\mu}$ is constant.
Note that this component vanishes if the gauge group $\cg$ is SU($N$)
but can be nonzero if $\cg$ contains an abelian component.
At mass dimension two,
the only contribution to the gauge-invariant action
is of the form
$\cl^{(2)} = - k^{(2)\mu\nu} \tr(F_{\mu\nu})$.
This operator also vanishes for SU($N$)
and more generally is a total derivative representing a surface term
when $k^{(2)\mu\nu}$ is constant,
so we omit it from Table \ref{tab:pure-gauge}.
At mass dimension three,
a gauge-invariant term of the form
$\cl^{(3)} \supset - k^{(3)\al\mu\nu} \tr(D_\al F_{\mu\nu})$
can be envisaged.
This is again a term vanishing for SU($N$)
and generically reducing to a total derivative 
when $k^{(3)\al\mu\nu}$ is constant,
so we omit it from the table as well.
Another contribution at mass dimension three 
contains the nonabelian Chern-Simon operator
shown in the table,
which is gauge invariant up to a surface term
when the corresponding coefficient $k^{(3)\ka}$ is constant.

For mass dimensions $d\geq 4$,
the construction \rf{puregauge} can be applied systematically.
As before,
factors of $\tr(F_{\mu\nu})$ and $\tr(D_\al F_{\mu\nu})$
vanish for SU($N$) and hence can be disregarded.
Note that the piece of the coefficient $k^{(4)\ka\la\mu\nu}$ 
proportional to products of the Minkowski metric is Lorentz invariant 
and therefore acts merely to renormalize 
the conventional Yang-Mills term $\cl_{A0}$.
The double trace $k^{(4)\ka\la}{}_{\ka\la}$ can therefore be set to zero
without loss of generality. 
Inspection of $\cl_A^{(8)}$ might suggest an additional term of the form
$\tr (F_{\ka\la}F_{\mu\nu}D_{(\al}D_{\be)} F_{\rh\si})$,
but up to surface terms it can be written as a linear combination
of the operators $\tr [ F_{\ka\la}(D_\al F_{\mu\nu})(D_\be F_{\rh\si}) ]$.

We note in passing that Table \ref{tab:pure-gauge}
is constructed assuming constant coefficients,
but allowing for spacetime dependence of the coefficients 
has minimal effect on the structure of the Lagrange density.
The terms shown in the table then appear 
with spacetime-dependent coefficients,
while any components of the Lagrange density
that are surface terms when the coefficients are constant 
become equivalent to existing terms at lower $d$
involving derivatives of the coefficients.
An effective reduction in mass dimension also occurs
when the gauge fields are constant backgrounds for the physics
\cite{dk16}.

The subset of $\cl_A$ consisting of terms quadratic in $A_\mu$
governs the propagation of the nonabelian gauge field.
Introducing the linearized nonabelian field strength 
$\fl_{\mu\nu} \equiv \prt_\mu A_\nu -\prt_\nu A_\mu$,
the quadratic terms for $d\geq 4$
are found to take the generic form
$ k^{\ka\la\mu\nu\cdots} 
\tr ( \fl_{\ka\la} \prt_{(n)} \fl_{\mu\nu} )$,
where the ellipsis on each coefficient $k^{\ka\la\mu\nu\cdots}$
is understood to be contracted with the $n$ indices on the derivatives.
Previous work 
\cite{km}
has shown that the quadratic Lagrange density in the QED limit 
can be written as a sum of two kinds of terms,
$k^{\ka\la\mu\nu\cdots} F_{\ka\la} \prt_{(n)} F_{\mu\nu}$ 
for even $n$
and $k^{\ka\cdots} \ep_{\ka\la\mu\nu} A^\la \prtn F^{\mu\nu}$ 
for odd $n$.
This matches the present result for even $n$,
while suggesting a correspondence of the form
\beq
k^{\ka\la\mu\nu\cdots} 
\tr ( \fl_{\ka\la} \prt_{(n)} \fl_{\mu\nu} )
\longleftrightarrow
k^{\al\cdots} \ep_{\al\la\mu\nu} 
\tr ( A^\la \prt_{(n+1)} F^{\textrm{L}\mu\nu} )
\label{corr}
\eeq
for odd $n$.

The correspondence \rf{corr} can be verified directly, 
assuming all coefficients are constants.
Up to surface terms,
the operator on the left-hand side of Eq.\ \rf{corr}
can be written as 
\beq
L_{\ka\la\mu\nu} =
\tr( -A_\la \prt_\ka \prtn \fl_{\mu\nu}+ A_\ka \prt_\la \prtn \fl_{\mu\nu}).
\eeq
Expanding $\prt_{(n+1)} = \prtn\prt_\ka$
and noting the antisymmetrization due to the Levi-Civita tensor,
the operator on the right-hand side of Eq.\ \rf{corr}
takes the form
\beq
R_{\ka\la\mu\nu} =
\tr( A_\la \prtn\prt_\ka \fl_{\mu\nu} + A_\mu \prtn \prt_\ka \fl_{\nu\la} + A_\nu \prtn\prt_\ka \fl_{\la\mu}).
\eeq
Direct calculation reveals that
$L_{\ka\la\mu\nu}$ and $R_{\ka\la\mu\nu}$
are linearly related up to surface terms,
\bea
&&
L_{\ka\la\mu\nu}+L_{\ka\mu\nu\la}+L_{\ka\nu\la\mu}=-R_{\ka\la\mu\nu}, 
\nn\\
&&
R_{\ka\la\mu\nu}-R_{\la\ka\mu\nu}=-2 L_{\ka\la\mu\nu},
\eea
thereby confirming the correspondence \rf{corr}.

The correspondence \rf{corr} reveals that for odd $d$ the operator
$\tr ( \fl_{\ka\la} \prt_{(n)} \fl_{\mu\nu} )$
can be viewed as a higher-$d$ generalization 
of the quadratic part of the Chern-Simons operator 
appearing in $\cL_g^{(3)}$.
By direct calculation modulo surface terms,
we find that the full nonlinear operator
$\tr(  F_{\ka\la} \Dnn F_{\mu\nu})$ 
with odd $d$ can be related to a generalized Chern-Simons operator
according to
\bea
k^{\ka\la\mu\nu\cdots} \tr( F_{\ka\la} \Dnn F_{\mu\nu})
&=&
-k_{CS}^{\ka\al\cdots} \ep_{\al\la\mu\nu} \tr( F_\ka^{\ \la} \Dnn F^{\mu\nu})
\nn\\
&&
\hskip -100pt
=k_{CS}^{\ka\al\cdots} \ep_{\al\la\mu\nu} 
\tr\big(
A^\la D_\ka \Dnn  F^{\mu\nu}
+ig[A^\la, A_\ka] \Dnn F^{\mu\nu}
\nn\\
&&
\hskip -50pt
+A_\ka \Dnn D^\la F^{\mu\nu}
-A_\ka D^\la \Dnn F^{\mu\nu}
\big),
\label{nonabcs}
\eea
where
\beq
k_{CS}^{\ka\al\cdots} = 
\half\ep^{\al}{}_{\la\mu\nu} k^{\ka\la\mu\nu\cdots}
\eeq
Together with the correspondence \rf{corr},
the result \rf{nonabcs} shows that for odd $d>3$ 
all terms in $\cl_A$ containing the linearized operators
$\ep^{\ka\la\mu\nu}\tr(A_\la \prt_{(n)} F^{L}_{\mu\nu})$
can be written as combinations of the operators
$\tr( F_{\ka\la} D_{(n-1)} F_{\mu\nu})$.
This provides strong support for the notion
that the form \rf{puregauge} incorporates every term with odd $d>3$
in the gauge-invariant action.
It remains an interesting open problem to prove this rigorously 
or to demonstrate a counterexample
by presenting a purely nonlinear gauge-invariant contribution 
to the action with odd $d>3$.

\subsection{Fermion-gauge sector}

The remaining part $\cl\pr$ of the full Lagrange density \rf{fullL}
is the fermion-gauge sector,
which involves products of operators appearing 
in the fermion and pure-gauge Lagrange densities
$\cl_\ps$ and $\cl_A$.
Suppressing indices,
the general structure of gauge-invariant terms
in the fermion-gauge sector can therefore be written schematically as
\bea
&&
\hskip -20pt
\cl\pr\supset 
k^{\cdots}\big( \tr \big[ (DF)\scdots(DF) \big] \big)
\scdots
\big( \tr \big[ (DF)\scdots(DF) \big] \big)
\nn\\
&&
\hskip 20pt 
\times 
[(\ov{D\ps})\Ga (DF)\scdots(DF)(D\ps)]
\scdots
\nn\\
&&
\hskip 30pt 
\times \scdots
[(\ov{D\ps})\Ga (DF)\scdots(DF)(D\ps)]
+\hc
\label{gfop}
\eea
These gauge-fermion terms typically appear
only at comparatively high values of the mass dimension $d$.
Consider,
for example,
gauge-fermion terms in an SU($N$) gauge theory.
The nontrivial gauge-invariant pure-gauge operator of lowest mass dimension
has structure $\tr(FF)$ 
because both $\tr(F)$ and $\tr(DF)$ vanish.
Also,
the gauge-invariant fermion-sector operators of lowest mass dimension
have structure $\psb\Ga\ps$.
The operator of lowest mass dimension in the gauge-fermion sector
therefore has the product structure $\tr(FF)\psb\Ga\ps$
and has $d=7$.

\renewcommand\arraystretch{1.4}
\begin{table*}
\caption{
\label{tab:QED}
Terms of low mass dimension $d\leq 6$
in the Lagrange density for single-fermion QED.}
\setlength{\tabcolsep}{8pt}
\begin{tabular}{cl}
\hline
\hline
Component & Expression \\
\hline
$	\cL_{0}	$	&		$		\half \psb (\ga^\mu i \prt_\mu - m_\ps ) \ps	+	\hc			
						-	\quar F_{\mu\nu}F^{\mu\nu}			$	\\	[2pt]
$	\cL^{(3)}_\ps	$	&		$	-	i m_5 \psb \ga_5\ps -a^\mu\psb\ga_\mu\ps					
						-	b^\mu\psb\gamma_5\ga_\mu\ps					
						-	\half H^{\mu\nu}\psb\simn \ps			$	\\	[2pt]
$	\cL^{(4)}_\ps	$	&		$		\half c^{\mu\al}\psb\ga_\mu i D_\al \ps					
						+	\half d^{\mu\al}\psb \ga_5 \ga_\mu i D_\al \ps					
						+	\half e^\al \psb iD_\al \ps					
						+	\half i f^\al \psb \ga_5 iD_\al \ps					
						+	\quar g^{\mu\nu\al} \psb \simn iD_\al \ps	+	\hc	$	\\	[2pt]
$	\cL_{\ps D}^{(5)}	$	&		$	-	\half m^{(5)\al\be}\psb iD_{(\al} iD_{\be)}\ps					
						-	\half i m_5^{(5)\al\be} \psb \ga_5 iD_{(\al} iD_{\be)} \ps			$	\\	[2pt]
			&	\hskip20pt	$	-	\half a^{(5)\mu\al\be} \psb \ga_\mu iD_{(\al} iD_{\be)} \ps					
						-	\half b^{(5)\mu\al\be}\psb \ga_5 \ga_\mu iD_{(\al} iD_{\be)} \ps					
						-	\quar H^{(5)\mu\nu\al\be} \psb \simn iD_{(\al} iD_{\be)} \ps	+	\hc	$	\\	[2pt]
$	\cL_{\ps F}^{(5)}	$	&		$	-	\half m_F^{(5)\al\be}  F_{\al\be} \psb \ps					
						-	\half im_{5F}^{(5)\al\be} F_{\al\be} \psb \ga_5 \ps					
						-	\half a_F^{(5)\mu\al\be} F_{\al\be} \psb \ga_\mu \ps					
						-	\half b_F^{(5)\mu\al\be} F_{\al\be} \psb \ga_5 \ga_\mu \ps					
						-	\quar H_F^{(5)\mu\nu\al\be} F_{\al\be} \psb \simn \ps			$	\\	[2pt]
$	\cL_{\ps D}^{(6)}	$	&		$		\half c^{(6)\mu\al\be\ga} \psb \ga_\mu iD_{(\al} iD_\be iD_{\ga)} \ps					
						+	\half d^{(6)\mu\al\be\ga} \psb \ga_5 \ga_\mu iD_{(\al} iD_\be iD_{\ga)} \ps			$	\\	[2pt]
			&	\hskip20pt	$	+	\half e^{(6)\al\be\ga} \psb iD_{(\al} iD_\be iD_{\ga)} \ps					
						+	\half i f^{(6)\al\be\ga} \psb \ga_5 iD_{(\al} iD_\be iD_{\ga)} \ps					
						+	\quar g^{(6)\mu\nu\al\be\ga} \psb \simn iD_{(\al} iD_\be iD_{\ga)} \ps	+	\hc	$	\\	[2pt]
$	\cL_{\ps F}^{(6)}	$	&		$		\quar c_F^{(6)\mu\al\be\ga} F_{\be\ga} \psb \ga_\mu iD_\al \ps					
						+	\quar d_F^{(6)\mu\al\be\ga} F_{\be\ga}\psb \ga_5 \ga_\mu iD_\al \ps			$	\\	[2pt]
			&	\hskip20pt	$	+	\quar e_F^{(6)\al\be\ga} F_{\be\ga}\psb iD_\al \ps					
						+	\quar i f_F^{(6)\al\be\ga}F_{\be\ga} \psb \ga_5 iD_\al \ps					
						+	\eigh g_F^{(6)\mu\nu\al\be\ga} F_{\be\ga}\psb \simn iD_\al \ps	+	\hc	$	\\	[2pt]
$	\cL_{\ps \prt F}^{(6)}	$	&		$	-	\half m_{\prt F}^{(6)\al\be\ga} (\prt_\al F_{\be\ga}) \psb \ps					
						-	\half i m_{5\prt F}^{(6)\al\be\ga} (\prt_\al F_{\be\ga}) \psb \ga_5 \ps			$	\\	[2pt]
			&	\hskip20pt	$	-	\half a_{\prt F}^{(6)\mu\al\be\ga} (\prt_\al F_{\be\ga})\psb \ga_\mu \ps					
						-	\half b_{\prt F}^{(6)\mu\al\be\ga}(\prt_\al F_{\be\ga}) \psb \ga_5 \ga_\mu \ps					
						-	\quar H_{\prt F}^{(6)\mu\nu\al\be\ga}  (\prt_\al F_{\be\ga})\psb \simn\ps			$	\\	[2pt]
$	\cL_{\ps\ps}^{(6)}	$	&		$		k_{SS}(\psb\ps)(\psb\ps)					
						-	k_{PP}(\psb\ga_5\ps)(\psb\ga_5\ps)					
						+	ik_{SP}(\psb\ps)(\psb\ga_5\ps)			$	\\	[2pt]
			&	\hskip20pt	$	+	(k_{SV})^\mu(\psb\ps)(\psb\ga_\mu\ps)					
						+	(k_{SA})^\mu (\psb\ps) (\psb\ga_5\ga_\mu\ps)					
						+	\half (k_{ST})^{\mu\nu}(\psb\ps)(\psb\simn\ps)			$	\\	[2pt]
			&	\hskip20pt	$	+	i(k_{PV})^\mu (\psb\ga_5\ps) (\psb\ga_\mu\ps)					
						+	i(k_{PA})^\mu (\psb\ga_5\ps) (\psb\ga_5\ga_\mu\ps)					
						+	\half i(k_{PT})^{\mu\nu} (\psb\ga_5\ps) (\psb\simn\ps)			$	\\	[2pt]
			&	\hskip20pt	$	+	\half (k_{VV})^{\mu \nu}(\psb\ga_\mu\ps)(\psb\ga_\nu\ps)					
						+	\half (k_{AA})^{\mu \nu}(\psb\ga_5\ga_\mu\ps)(\psb\ga_5\ga_\nu\ps)					
						+	(k_{V\!A})^{\mu\nu}(\psb\ga_\mu\ps) (\psb \ga_5\ga_\mu \ps)			$	\\	[2pt]
			&	\hskip20pt	$	+	\half (k_{VT})^{\la\mu\nu} (\psb \ga_\la \ps) (\psb \simn \ps)					
						+	\half (k_{AT})^{\la\mu\nu} (\psb \ga_5\ga_\la \ps) (\psb \simn \ps)					
						+	\eigh (k_{TT})^{\ka\la\mu\nu}(\psb\si_{\ka\la}\ps)(\psb\simn\ps)			$	\\	[2pt]
$	\cL_A^{(1)}	$	&		$	-	(k_A)^{\mu} A_\mu			$	\\	[2pt]
$	\cL_A^{(3)}	$	&		$		\half (k_{AF})^{\ka} \ep_{\ka\la\mu\nu} A^\la F^{\mu\nu}			$	\\	[2pt]
$	\cL_A^{(4)}	$	&		$	-	\quar (k_F)^{\ka\la\mu\nu} F_{\ka\la} F_{\mu\nu}			$	\\	[2pt]
$	\cL_A^{(5)}	$	&		$	-	\quar k^{(5)\al\ka\la\mu\nu} F_{\ka\la} \prt_\al F_{\mu\nu}			$	\\	[2pt]
$	\cL_A^{(6)}	$	&		$	-	\quar k_\prt^{(6)\al\be\ka\la\mu\nu} F_{\ka\la} \prt_{\al} \prt_{\be} F_{\mu\nu}					
						-	\tfrac 1 {12} k_F^{(6)\ka\la\mu\nu\rh\si}  F_{\ka\la} F_{\mu\nu} F_{\rh\si}			$	\\	[2pt]
\hline
\hline
\end{tabular}
\end{table*}

The gauge-fermion operators of the form \rf{gfop}
are relevant only for the nonabelian case
because for U(1) all operators are already included 
in the construction of the fermion sector.
For example,
the abelian versions of terms of the form $\tr(F)\psb\Ga\ps$ 
are included as part of $\cL_{\ps F}^{(5)}$
in Table \ref{tab:spinor},
while ones of the form $\tr(DF)\psb\Ga\ps$ 
are included in $\cL_{\ps DF}^{(6)}$.

Since the SM gauge group 
is SU(3)$\times$SU(2)$\times$U(1),
no terms in the fermion-gauge sector appear for $d\leq 6$.
We can therefore disregard $\cl\pr$
in physical applications of the SME
involving operators with these comparatively low mass dimensions.
The same line of reasoning holds
for the gauge theory combining QCD and QED,
as the corresponding gauge group is
SU(3)$\times$U(1).
However,
more general gauge theories
or investigations of physical effects in fermion-gauge couplings
for $d>6$ may require inclusion of these extra gauge-invariant terms.

\subsection{Limiting cases}
\label{sec:limits}

For certain specific applications,
it can be useful to consider restrictions
of the general Lagrange density \rf{fullL}
to a subset of relevant terms.
In this section,
we discuss in turn the limiting cases
of Lorentz-violating QED,
Lorentz-violating QCD and QED, 
Lorentz-invariant gauge field theories,
and isotropic Lorentz-violating models.

\subsubsection{Lorentz-violating QED}
\label{sec:QED}

One limit with broad applicability 
is the abelian restriction of the general nonabelian gauge theory
obtained by choosing the gauge group $\cg$ to be U(1).
With the Dirac field identified as the electron field
and the gauge field identified with the photon,
we obtain a Lorentz- and CPT-violating extension 
of the conventional theory of QED for electrons and photons.
The abelian restriction is also relevant
for the description of electromagnetic interactions
of other fermions,
including ones that are uncharged.

A generic term in the Lagrange density 
for the abelian theory takes the form
\beq
\cl\supset
k^{\cdots}(\prt F) \cdots (\prt F)
(\ov{D\ps})\Ga (D\ps) \cdots ( \ov{D\ps})\Ga (D\ps)+\hc ,
\label{qedterm}
\eeq
where all spacetime indices and gamma matrices are understood,
$\prt$ represents $\prtn$,
and $D$ represents $\Dnn$.
Note that the factors involving the gauge field strength
decouple from the fermion in gauge space,
and hence the term \rf{qedterm}
can be treated as the product of distinct factors
involving either the field strength or the fermion fields.
To avoid redundancy in the term \rf{qedterm}, 
the basis \rf{abelian} for the gauge-covariant operators
can be chosen.

Table \ref{tab:QED} 
presents all terms with mass dimension six or less
in the Lagrange density for the single-fermion QED limit.
The first column displays the components of the Lagrange density,
while the second column provides
the explicit form of the terms in each component.
The component $\cl_0$ is the conventional Lagrange density for QED. 
Other components with $d\leq 4$  
are power-counting renormalizable
and form the minimal QED extension
\cite{ck}.
The coefficients for these minimal terms are denoted
using conventions widely adopted in the literature.
The components of the Lagrange density with $d\geq 5$
in the fermion sector are separated into pieces
according to the scheme adopted in Table \ref{tab:spinor}.
Note that both Lorentz-invariant and Lorentz-violating terms 
are encompassed by the entries in the table,
depending on the structure of the coefficients.
Operators with odd numbers of spacetime indices
are odd under CPT transformations,
and they all break Lorentz symmetry
\cite{ck,owg}.

The coefficients of all terms in the table
have dimension GeV$^{4-d}$ and are real.
They can be assumed constant in an inertial frame near the Earth
\cite{ck,ak04}.
In the fermion sector
the spacetime indices $\mu$, $\nu$ are linked
to the gamma matrices and hence to spin properties,
while in the pure-gauge sector they are associated
with the gauge field strengths. 
In both sectors,
the spacetime indices $\al$, $\be$, $\ldots$ 
are associated with derivatives. 
As before,
parentheses around $n$ spacetime indices 
represent total symmetrization with a factor of $1/n!$.

Where possible,
the table follows the notation for the fermion sector 
adopted in Ref.\ \cite{dk16}.
Also,
all quadratic terms in the pure-gauge sector of the QED extension
have been obtained in Ref.\ \cite{km}.
In contrast,
the terms representing self interactions,
including those in $\cL_{\ps\ps}^{(6)}$ 
and the one cubic in $F$ in $\cL_{A}^{(6)}$,
represent new arenas for investigation.

Table \ref{tab:pure-gauge} contains also contributions
to the nonabelian pure-gauge action for $d=7$ and 8.
For operators of mass dimension seven,
restricting these terms to the QED limit 
yields the expressions
\bea
\cL_A^{(7)}
&=&
-\quar k_\prt^{(7)\al\be\ga\ka\la\mu\nu} 
F_{\ka\la} \prt_{\al} \prt_\be \prt_{\ga} F_{\mu\nu}
\nn\\
&&
-\tfrac 1{12} k_F^{(7)\al\ka\la\mu\nu\rh\si} 
F_{\ka\la} F_{\mu\nu} \prt_\al F_{\rh\si},
\eea
while for those of mass dimension eight we find
\bea
\cL_A^{(8)}
&=&
-\quar k_\prt^{(8)\al\be\ga\de\ka\la\mu\nu} 
F_{\ka\la} \prt_{\al} \prt_\be \prt_\ga \prt_{\de} F_{\mu\nu}
\nn\\
&&
-\tfrac 1 {12} k_{\prt F}^{(8)\al\be\ka\la\mu\nu\rh\si} 
F_{\ka\la}(\prt_\al F_{\mu\nu})(\prt_\be F_{\rh\si})
\nn\\
&&
-\tfrac 1 {48} k_F^{(8) \ka\la\mu\nu\rh\si\ta\up} 
F_{\ka\la}F_{\mu\nu}F_{\rh\si}F_{\ta\up}.
\label{deight}
\eea
The nonlinear interactions of the photon predicted by these terms 
are also of potential interest
in both the theoretical and experimental contexts.
We revisit this point in Sec.\ \ref{Lightbylight} below.

\renewcommand\arraystretch{1.6}
\begin{table*}
\caption{
\label{tab:QCD}
Terms of low mass dimension $d\leq 6$
in the Lagrange density for QCD and QED with multiple flavors of quarks.}
\setlength{\tabcolsep}{8pt}
\begin{tabular}{cl}
\hline
\hline
Component & Expression \\
\hline
$	\cL_{0}	$	&		$		\half \psb_A \ga^\mu i D_\mu \ps_A	+	\hc			
						-	\psb_A m_{AB} \ps_B					
						-	\quar F_{\mu\nu}F^{\mu\nu}					
						-	\half \tr(G_{\mu\nu}G^{\mu\nu})			$	\\	[2pt]
$	\cL^{(3)}_\ps	$	&		$	-	i m_{5AB} \psb_A \ga_5\ps_B -a^\mu_{AB} \psb_A\ga_\mu\ps_B					
						-	b^\mu_{AB} \psb_A\gamma_5\ga_\mu\ps_B					
						-	\half H^{\mu\nu}_{AB}\psb_A\simn \ps_B			$	\\	[2pt]
$	\cL^{(4)}_\ps	$	&		$		\half c^{\mu\al}_{AB}\psb_A\ga_\mu i D_\al \ps_B					
						+	\half d^{\mu\al}_{AB}\psb_A \ga_5 \ga_\mu i D_\al \ps_B			$	\\	[2pt]
			&	\hskip20pt	$	+	\half e^\al_{AB} \psb_A iD_\al \ps_B					
						+	\half i f^\al_{AB} \psb_A \ga_5 iD_\al \ps_B					
						+	\quar g^{\mu\nu\al}_{AB} \psb_A \simn iD_\al \ps_B	+	\hc	$	\\	[2pt]
$	\cL_{\ps D}^{(5)}	$	&		$	-	\half (m^{(5)})^{\al\be}_{AB}\psb_A iD_{(\al} iD_{\be)}\ps_B					
						-	\half i (m_5^{(5)})_{AB}^{\al\be} \psb_A \ga_5 iD_{(\al} iD_{\be)} \ps_B					
						-	\half (a^{(5)})^{\mu\al\be}_{AB} \psb_A \ga_\mu iD_{(\al} iD_{\be)} \ps_B			$	\\	[2pt]
			&	\hskip20pt	$	-	\half (b^{(5)})^{\mu\al\be}_{AB} \psb_A \ga_5 \ga_\mu iD_{(\al} iD_{\be)} \ps_B					
						-	\quar (H^{(5)})^{\mu\nu\al\be}_{AB} \psb_A \simn iD_{(\al} iD_{\be)} \ps_B	+	\hc	$	\\	[2pt]
$	\cL_{\ps F}^{(5)}	$	&		$	-	\half (m_F^{(5)})_{AB}^{\al\be} F_{\al\be} \psb_A \ps_B					
						-	\half i(m_{5F}^{(5)})_{AB}^{\al\be} F_{\al\be} \psb_A \ga_5 \ps_B			$	\\	[2pt]
			&	\hskip20pt	$	-	\half (a_F^{(5)})_{AB}^{\mu\al\be} F_{\al\be} \psb_A \ga_\mu \ps_B					
						-	\half (b_F^{(5)})_{AB}^{\mu\al\be} F_{\al\be} \psb_A \ga_5 \ga_\mu \ps_B					
						-	\quar (H_F^{(5)})_{AB}^{\mu\nu\al\be} F_{\al\be} \psb_A \simn \ps_B			$	\\	[2pt]
$	\cL_{\ps G}^{(5)}	$	&		$	-	\half (m_G^{(5)})_{AB}^{\al\be} \psb_A G_{\al\be} \ps_B					
						-	\half i(m_{5G}^{(5)})_{AB}^{\al\be} \psb_A \ga_5 G_{\al\be} \ps_B			$	\\	[2pt]
			&	\hskip20pt	$	-	\half (a_G^{(5)})_{AB}^{\mu\al\be} \psb_A \ga_\mu G_{\al\be} \ps_B					
						-	\half (b_G^{(5)})_{AB}^{\mu\al\be} \psb_A \ga_5 \ga_\mu G_{\al\be} \ps_B					
						-	\quar (H_G^{(5)})_{AB}^{\mu\nu\al\be} \psb_A \simn G_{\al\be} \ps_B			$	\\	[2pt]
$	\cL_{\ps D}^{(6)}	$	&		$		\half (c^{(6)})^{\mu\al\be\ga}_{AB} \psb_A \ga_\mu iD_{(\al} iD_\be iD_{\ga)} \ps_B					
						+	\half (d^{(6)})^{\mu\al\be\ga}_{AB} \psb_A \ga_5 \ga_\mu iD_{(\al} iD_\be iD_{\ga)} \ps_B					
						+	\half (e^{(6)})^{\al\be\ga}_{AB} \psb_A iD_{(\al} iD_\be iD_{\ga)} \ps_B			$	\\	[2pt]
			&	\hskip20pt	$	+	\half i (f^{(6)})^{\al\be\ga}_{AB} \psb_A \ga_5 iD_{(\al} iD_\be iD_{\ga)} \ps_B					
						+	\quar (g^{(6)})^{\mu\nu\al\be\ga}_{AB} \psb_A \simn iD_{(\al} iD_\be iD_{\ga)} \ps_B	+	\hc	$	\\	[2pt]
$	\cL_{\ps F}^{(6)}	$	&		$		\quar (c_F^{(6)})_{AB}^{\mu\al\be\ga} F_{\be\ga} \psb_A \ga_\mu iD_\al \ps_B					
						+	\quar (d_F^{(6)})_{AB}^{\mu\al\be\ga} F_{\be\ga}\psb_A \ga_5 \ga_\mu iD_\al \ps_B			$	\\	[2pt]
			&	\hskip20pt	$	+	\quar (e_F^{(6)})_{AB}^{\al\be\ga} F_{\be\ga}\psb_A iD_\al \ps_B					
						+	\quar i (f_F^{(6)})_{AB}^{\al\be\ga}F_{\be\ga} \psb_A \ga_5 iD_\al \ps_B					
						+	\eigh (g_F^{(6)})_{AB}^{\mu\nu\al\be\ga} F_{\be\ga}\psb_A \simn iD_\al \ps_B	+	\hc	$	\\	[2pt]
$	\cL_{\ps G}^{(6)}	$	&		$		\quar (c_G^{(6)})_{AB}^{\mu\al\be\ga} \psb_A \ga_\mu G_{\be\ga} iD_\al \ps_B					
						+	\quar (d_G^{(6)})_{AB}^{\mu\al\be\ga} \psb_A \ga_5 \ga_\mu G_{\be\ga} iD_\al \ps_B			$	\\	[2pt]
			&	\hskip20pt	$	+	\quar (e_G^{(6)})_{AB}^{\al\be\ga} \psb_A G_{\be\ga} iD_\al \ps_B					
						+	\quar i (f_G^{(6)})_{AB}^{\al\be\ga} \psb_A \ga_5 G_{\be\ga} iD_\al \ps_B					
						+	\eigh (g_G^{(6)})_{AB}^{\mu\nu\al\be\ga} \psb_A \simn G_{\be\ga} iD_\al \ps_B	+	\hc	$	\\	[2pt]
$	\cL_{\ps \prt F}^{(6)}	$	&		$	-	\half (m_{\prt F}^{(6)})_{AB}^{\al\be\ga} (\prt_\al F_{\be\ga}) \psb_A \ps_B					
						-	\half i (m_{5\prt F}^{(6)})_{AB}^{\al\be\ga} (\prt_\al F_{\be\ga}) \psb_A \ga_5 \ps_B			$	\\	[2pt]
			&	\hskip20pt	$	-	\half (a_{\prt F}^{(6)})_{AB}^{\mu\al\be\ga} (\prt_\al F_{\be\ga})\psb_A \ga_\mu \ps_B					
						-	\half (b_{\prt F}^{(6)})_{AB}^{\mu\al\be\ga}(\prt_\al F_{\be\ga}) \psb_A \ga_5 \ga_\mu \ps_B					
						-	\quar (H_{\prt F}^{(6)})_{AB}^{\mu\nu\al\be\ga} (\prt_\al F_{\be\ga})\psb_A \simn\ps_B			$	\\	[2pt]
$	\cL_{\ps DG}^{(6)}	$	&		$	-	\half (m_{DG}^{(6)})_{AB}^{\al\be\ga} \psb_A (D_\al G_{\be\ga})\ps_B					
						-	\half i (m_{5DG}^{(6)})_{AB}^{\al\be\ga} \psb_A \ga_5 (D_\al G_{\be\ga})\ps_B			$	\\	[2pt]
			&	\hskip20pt	$	-	\half (a_{DG}^{(6)})_{AB}^{\mu\al\be\ga} \psb_A \ga_\mu (D_\al G_{\be\ga})\ps_B					
						-	\half (b_{DG}^{(6)})_{AB}^{\mu\al\be\ga} \psb_A \ga_5 \ga_\mu (D_\al G_{\be\ga})\ps_B					
						-	\quar (H_{DG}^{(6)})_{AB}^{\mu\nu\al\be\ga} \psb_A \simn (D_\al G_{\be\ga})\ps_B			$	\\	[2pt]
$	\cL_{\ps\ps}^{(6)}	$	&		$		(k_{SS})_{ABCD}(\psb_A\ps_B)(\psb_C \ps_D)					
						-	(k_{PP})_{ABCD}(\psb_A\ga_5\ps_B)(\psb_C\ga_5\ps_D)			$	\\	[2pt]
			&	\hskip20pt	$	+	i(k_{SP})_{ABCD}(\psb_A\ps_B)(\psb_C\ga_5\ps_D)					
						+	(k_{SV})_{ABCD}^\mu(\psb_A\ps_B)(\psb_C\ga_\mu\ps_D)			$	\\	[2pt]
			&	\hskip20pt	$	+	(k_{SA})_{ABCD}^\mu (\psb_A\ps_B) (\psb_C\ga_5\ga_\mu\ps_D)					
						+	\half (k_{ST})_{ABCD}^{\mu\nu}(\psb_A\ps_B)(\psb_C\simn\ps_D)			$	\\	[2pt]
			&	\hskip20pt	$	+	i(k_{PV})_{ABCD}^\mu (\psb_A\ga_5\ps_B) (\psb_C\ga_\mu\ps_D)					
						+	i(k_{PA})_{ABCD}^\mu (\psb_A\ga_5\ps_B) (\psb_C\ga_5\ga_\mu\ps_D)			$	\\	[2pt]
			&	\hskip20pt	$	+	\half i(k_{PT})_{ABCD}^{\mu\nu} (\psb_A\ga_5\ps_B) (\psb_C\simn\ps_D)					
						+	\half (k_{VV})_{ABCD}^{\mu \nu}(\psb_A\ga_\mu\ps_B)(\psb_C\ga_\nu\ps_D)			$	\\	[2pt]
			&	\hskip20pt	$	+	\half (k_{AA})_{ABCD}^{\mu \nu}(\psb_A\ga_5\ga_\mu\ps_B)(\psb_C\ga_5\ga_\nu\ps_D)					
						+	(k_{V\!A})_{ABCD}^{\mu\nu}(\psb_A\ga_\mu\ps_B) (\psb_C \ga_5\ga_\mu \ps_D)			$	\\	[2pt]
			&	\hskip20pt	$	+	\half (k_{VT})_{ABCD}^{\la\mu\nu} (\psb_A \ga_\la \ps_B) (\psb_C \simn \ps_D)					
						+	\half (k_{AT})_{ABCD}^{\la\mu\nu} (\psb_A \ga_5\ga_\la \ps_B) (\psb_C\simn \ps_D)			$	\\	[2pt]
			&	\hskip20pt	$	+	\eigh (k_{TT})_{ABCD}^{\ka\la\mu\nu}(\psb_A\si_{\ka\la}\ps_B)(\psb_C\simn\ps_D)			$	\\	[2pt]
$	\cL_{AG}^{(1)}	$	&		$	-	(k_A)^{\mu} A_\mu			$	\\	[2pt]
$	\cL_{AG}^{(3)}	$	&		$		\half (k_{AF})^{\ka} \ep_{\ka\la\mu\nu} A^\la F^{\mu\nu}					
						+	(k_3)^{\ka} \ep_{\ka\la\mu\nu} \tr(G^\la G^{\mu\nu}+\tfrac23 ig_3 G^\la G^\mu G^\nu)			$	\\	[2pt]
$	\cL_{AG}^{(4)}	$	&		$	-	\quar (k_F)^{\ka\la\mu\nu} F_{\ka\la} F_{\mu\nu}					
						-	\half (k_G)^{\ka\la\mu\nu}\tr(G_{\ka\la}G_{\mu\nu})			$	\\	[2pt]
$	\cL_{AG}^{(5)}	$	&		$	-	\quar k^{(5)\al\ka\la\mu\nu} F_{\ka\la} \prt_\al F_{\mu\nu}					
						-	\half k_D^{(5)\al\ka\la\mu\nu}\tr(G_{\ka\la}D_\al G_{\mu\nu})			$	\\	[2pt]
$	\cL_{AG}^{(6)}	$	&		$	-	\quar k_\prt^{(6)\al\be\ka\la\mu\nu} F_{\ka\la} \prt_{\al} \prt_{\be} F_{\mu\nu}					
						-	\tfrac 1 {12} k_F^{(6)\ka\la\mu\nu\rh\si}  F_{\ka\la} F_{\mu\nu} F_{\rh\si}			$	\\	[2pt]
			&	\hskip20pt	$	-	\half k_D^{(6)\al\be\ka\la\mu\nu}\tr(G_{\ka\la}D_{(\al}D_{\be)}G_{\mu\nu})					
						-	\tfrac16 k_G^{(6)\ka\la\mu\nu\rh\si}\tr(G_{\ka\la}G_{\mu\nu}G_{\rh\si})					
						-	\quar k_{FG}^{(6)\ka\la\mu\nu\rh\si}F_{\ka\la}\tr(G_{\mu\nu}G_{\rh\si})			$	\\	[2pt]
\hline
\end{tabular}
\end{table*}

\subsubsection{Lorentz-violating QCD and QED}

Another interesting model contained in our general framework
is the limit of QCD and QED coupled to quarks.
For this model,
the gauge group $\cg$ is SU(3)$\times$U(1)
and we allow quark fields $\ps_A$ with multiple flavors
labeled by indices $A$, $B$, $\ldots$.
In the usual six-quark scenario,
these indices span the values $u$, $d$, $s$, $c$, $b$, and $t$.
Each quark lies in the {\bf 3} representation of SU(3)
and carries U(1) charge denoted $q_A$.

The covariant derivative acting on the quark fields
can be written as
$D_\mu\ps_A=(\prt_\mu+iq_A A_\mu-ig_3 G_\mu)\ps_A$,
where $A_\mu$ is the photon field,
$g_3$ is the QCD coupling constant,
and $G_\mu$ is the gluon field.
Acting on the photon field strength $F_{\mu\nu}$,
the covariant derivative gives
$D_\al F_{\mu\nu}=\prt_\al F_{\mu\nu}$,
while acting on gluon field strength $G_{\mu\nu}$ gives 
$D_\al G_{\mu\nu}=\prt_\al G_{\mu\nu}-ig_3 [G_\al, G_{\mu\nu}]$.

Table \ref{tab:QCD} lists all terms with mass dimension six or less
in the Lagrange density for this model.
The first column shows the components of the Lagrange density,
and the second column displays the explicit expression
for each component.
The conventional Lorentz-invariant Lagrange density $\cL_{0}$ 
for QCD and QED coupled to quarks
is given in the first line of the table.
The terms in the model with $d\leq 4$ 
are power-counting renormalizable
and form a subset of the minimal SME
\cite{ck,ak04}.
The notation for the corresponding coefficients
is chosen to match the standard one in the literature.
The table includes both Lorentz-invariant and Lorentz-violating terms.
All terms with an odd number of spacetime indices
are odd under CPT.

\renewcommand\arraystretch{1.4}
\begin{table*}
\caption{
\label{tab:LI}
Lorentz-invariant components of operators in gauge field theories.}
\setlength{\tabcolsep}{8pt}
\begin{tabular}{llll}
\hline
\hline
Theory & Operator & Dimension & Lorentz-invariant components  \\
\hline
nonabelian	&	$	\psb \Ga_I D_{(2n-2)} \psi	+	\hc	$	&	odd,	$	d=2n+1	$	&	$	\psb D^{\textrm{LI}}_{(2n-2)} \psi	+	\hc	,\			
														i\psb \ga_5 D^{\textrm{LI}}_{(2n-2)}\psi	+	\hc		$	\\	[2pt]
	&	$	\psb \Ga_I D_{(2n-1)} \psi	+	\hc	$	&	even,	$	d=2n+2	$	&	$	\psb\ga^\mu D_\mu D^{\textrm{LI}}_{(2n-2)} \psi	+	\hc	,\			
														\psb \ga_5\ga^\mu D_\mu D^{\textrm{LI}}_{(2n-2)}\psi	+	\hc		$	\\	[2pt]
	&	$	\psb \Ga_I F D_{(2n-2)}  \psi	+	\hc	$	&	odd,	$	d=2n+3	$	&	$	\half\psb \si^{\mu\nu}F_{\mu\nu} D^{\textrm{LI}}_{(2n-2)}\psi	+	\hc	,\			
														\half\psb \si^{\mu\nu}\Ftil_{\mu\nu} D^{\textrm{LI}}_{(2n-2)}\psi	+	\hc		$	\\	[2pt]
	&	$	\psb \Ga_I F D_{(2n-1)} \psi	+	\hc	$	&	even,	$	d=2n+4	$	&	$	\psb\ga^\mu F_{\mu\nu} D^\nu D^{\textrm{LI}}_{(2n-2)} \psi	+	\hc	,\			
														\psb\ga^\mu \Ftil_{\mu\nu} D^\nu D^{\textrm{LI}}_{(2n-2)} \psi	+	\hc	,\	$	\\	[2pt]
	&						&					&	$	\psb\ga_5\ga^\mu F_{\mu\nu} D^\nu D^{\textrm{LI}}_{(2n-2)} \psi	+	\hc	,\			
														\psb\ga_5\ga^\mu \Ftil_{\mu\nu} D^\nu D^{\textrm{LI}}_{(2n-2)} \psi	+	\hc		$	\\	[2pt]
nonabelian	&	$	\tr(FF\cdots F)			$	&	even, 	$	d=2n+2	$	&		traces of full contractions of products of $F_{\mu\nu}$, $\et^{\mu\nu}$, $\ep^{\ka\la\mu\nu}$					\\	[2pt]
	&	$	\tr(FD_{(2n)} F)			$	&	even, 	$	d=2n+4	$	&	$	\tr( F_{\mu\nu}D_{(2n)}^{\textrm{LI}} F^{\mu\nu})			,\			
														\tr( F_{\mu\nu}D_{(2n)}^{\textrm{LI}} \Ftil^{\mu\nu})				$	\\	[4pt]
abelian	&	$	FF\cdots F			$	&	even, 	$	d=4n	$	&	$	\PP^n, \PP^{n-1}\QQ, \PP^{n-2} \QQ^2, \cdots, \PP \QQ^{n-1}, \QQ^n				$	\\	[2pt]
	&	$	F\prt_{(2n)} F			$	&	even,	$	d=2n+4	$	&	$	F_{\mu\nu} (\prt^\al \prt_\al)^{n} F^{\mu\nu}				$	\\	[2pt]
\hline
\hline
\end{tabular}
\end{table*}

For an operator of mass dimension $d$,
the corresponding coefficient has mass dimension ${4-d}$.
The coefficients in the fermion sector
are tensors in flavor space with complex entries
restricted by the requirement that the Lagrange density is hermitian, 
while
the coefficients in the gauge sector are real.
When evaluated in an inertial frame near the Earth,
all coefficients can be treated as spacetime constants
\cite{ck,ak04}.
The notation for the spacetime indices
follows that for Table \ref{tab:QED}.

The model describes the general interactions 
of quarks, photons, and gluons,
and it can be viewed as an effective field theory
incorporating all Lorentz invariant terms
along with violations of Lorentz and CPT symmetry.
Note that the presence of two gauge groups in the model
implies that the pure-gauge sector contains terms
involving cross couplings of both sets of gauge fields.
For example,
the last term in $\cl^{(6)}_{AG}$ in the table
represents a photon-gluon-gluon interaction.
We remark that this term has a Lorentz-invariant component 
of potential phenomenological interest,
although Furry's theorem prevents its generation
via one-loop radiative corrections in the conventional SM.
Also,
the presence of multiple fermion flavors in the model
leads to flavor dependence of the coefficients in the fermion sector
and hence introduces various flavor-mixing effects.
Overall,
the model predicts a rich phenomenology
with many avenues open for experimental exploration.
One aspect of this is considered
in Sec.\ \ref{Deep inelastic scattering} below.

\subsubsection{Lorentz-invariant limit}
\label{Lorentz-invariant limit}

The general Lagrange density $\cL$ 
for the nonabelian gauge theory \rf{fullL}
incorporates both Lorentz-invariant and Lorentz-violating pieces.
A given term is Lorentz invariant 
if the coefficient can be written as a product
of the invariant tensors of the Lorentz group,
which are the Minkowski metric $\et^{\mu\nu}$
and the Levi-Civita tensor $\ep^{\ka\la\mu\nu}$.
As a simple example,
any part of the coefficient $c^{\mu\al}$ 
proportional to $\et^{\mu\al}$
generates a Lorentz-invariant contribution to $\cl^{(4)}_\ps$ 
in Table \ref{tab:spinor}.
All the Lorentz-invariant operators are isotropic in any inertial frame,
and they are also all CPT even.

Since both $\et^{\mu\nu}$ and $\ep^{\ka\la\mu\nu}$ 
have an even number of indices,
no coefficient with an odd number of indices
can be expressed as a Lorentz-invariant tensor.
Equivalently,
no operator with an odd number of spacetime indices
can contain a Lorentz-invariant piece.
In the fermion sector,
this implies only about half the terms in Table \ref{tab:spinor}
contain Lorentz-invariant components.
In the pure-gauge sector,
we find the elegant result that no Lorentz-invariant terms exist
for operators of odd mass dimension $d$.

Table \ref{tab:LI} displays the Lorentz-invariant components
of certain operators in gauge field theories.
The first column identifies the nonabelian and abelian cases.
Note that the abelian fermion sector 
is a direct limit of the nonabelian one.
Each entry in the second column 
lists a specific type of operator in schematic form,
with all spacetime indices omitted
and using the placeholder $\Ga_I$ for any gamma-matrix structures.
The third column shows the allowed mass dimension $d$ of each operator,
which is specified in terms of an integer $n=1,2,\ldots$.
The final column provides the possible Lorentz-invariant components.
In these expressions,
the covariant derivatives are understood to be totally symmetrized,
and we use the abbreviations
\beq
\PP=F_{\mu\nu}F^{\mu\nu},
\quad
\QQ=F_{\mu\nu}\Ftil^{\mu\nu},
\quad
D^{\textrm{LI}}_{(2n)}=(D_\mu D^\mu)^{n},
\label{pqd}
\eeq
where $\Ftil^{\mu\nu}=\ep^{\mu\nu\rh\si}F_{\rh\si}/2$ 
is the dual field strength.

In the QED limit,
all Lorentz-invariant terms in the pure-gauge sector
must be constructed as combinations of the two invariants $\PP$ and $\QQ$,
which can be expressed in terms 
of the electric field $\mathbf{E}$
and the magnetic induction $\mathbf{B}$
as 
$\PP=-2(\mathbf{E}^2-\mathbf{B}^2)$
and
$\QQ=-4\mathbf{E}\cdot \mathbf{B}$.
Since both $\PP$ and $\QQ$ have mass dimension four,
the mass dimensions of all Lorentz-invariant pure-gauge terms
must be multiples of four.
For $d=4n$,
the independent terms number $n+1$ 
and each is a monomial of order $n$ in $\PP$ and $\QQ$.
Note that for the special case $n=1$
the monomial $\QQ$ is a total derivative
and so can be disregarded in the Lagrange density
if its scalar coefficient is constant.
Related results hold in the full nonabelian theory.
However,
as $\PP$ and $\QQ$ then have a nontrivial commutator,
there are typically more independent Lorentz-invariant monomials 
in the nonabelian case.

The pure-gauge sector also contains quadratic terms
of the schematic form $\tr(F D_{(2n)} F)$,
as shown in the table.
The Lorentz-invariant components of these
can be obtained directly by contracting indices.
A useful identity in this respect is
$2F_{\mu\rh}D^{\mu} D_{\nu} D^{\textrm{LI}}_{(2n-2)} F^{\nu\rh}
=F_{\mu\rh} D^{\textrm{LI}}_{(2n)} F^{\mu\rh}$,
which can be proved by dualizing the field strengths
and using the homogeneous equation $D^\mu \Ftil_{\mu\si}=0$.
This procedure yields the two nonabelian quadratic pure-gauge terms 
presented in the table.
In the abelian limit,
the homogeneous equation implies 
that the second expression can be neglected up to a surface term,
thereby leaving only one Lorentz-invariant quadratic contribution at each $n$.

\renewcommand\arraystretch{1.4}
\begin{table*}
\caption{
\label{tab:isotropic}
Isotropic components of operators in gauge field theories.}
\setlength{\tabcolsep}{4pt}
\begin{tabular}{lllll}
\hline
\hline
Theory & Operator & Dimension & Isotropic components & Number \\
\hline
nonabelian	&	$	\psb \Ga_I \psi	+	\hc	$	&		$	d=	3	$	&	$	\psb\ps	+	\hc	,\							
															i\psb \ga_5 \ps	+	\hc	,\							
															\psb\ga_0 \ps	+	\hc	,\							
															\psb\ga_5\ga_0 \ps	+	\hc		$	&		4		\\	[2pt]
	&	$	\psb \Ga_I \Dnn \psi	+	\hc	$	&		$	d\geq	4	$	&	$	\psb\Dison\ps	+	\hc	,\							
															i\psb \ga_5 \Dison\ps	+	\hc		$	&				\\	[2pt]
	&						&						&	$	\psb\ga_0 \Dison\ps	+	\hc	,\							
															\psb\ga_5\ga_0 \Dison\ps	+	\hc		$	&				\\	[2pt]
	&						&						&	$	\psb\ga_j D^j\Diso_{(n-1)}\ps	+	\hc	,\							
															\psb\ga_5\ga_j D^j \Diso_{(n-1)}\ps	+	\hc		$	&				\\	[2pt]
	&						&						&	$	\half \psb\si_{0j}D^j\Diso_{(n-1)}\ps	+	\hc	,\							
															\quar \psb \ep_{jkl}\si^{kl}D^j\Diso_{(n-1)}\ps	+	\hc		$	&	$	$total$~ 4d-8	$	\\	[2pt]
nonabelian	&	$	\cL_A^{(3)}			$	&		$	d=	3	$	&	$	\ep_{jkl} \tr(A^j F^{kl}+\tfrac23igA^j A^k A^l)				$	&		1		\\	[2pt]
	&	$	\tr(FF\cdots F)			$	&		$	d=2n+2		$	&		traces of full contractions of products of $E^j$, $B^j$, $\de_{jk}$, and $\ep_{jkl}$					&				\\	[2pt]
	&	$	\tr(F D_{(n-1)} F)			$	&	even,	$	d\geq	4	$	&	$	\tr( F_{0j}\Diso_{(n-1)}F^{0j})			,\							
															\tr(F_{0j}\Diso_{(n-1)}\Ftil^{0j})			,\							
															\tr( \Ftil_{0j}\Diso_{(n-1)}\Ftil^{0j})				$	&	$	\tfrac 3 2 (d-2)	$	\\	[2pt]
	&	$	\tr(F\Dnn F)			$	&	odd,	$	d\geq	5	$	&	$	\tr(\ep_{jkl}F^{0j}D^k\Diso_{(n-1)}F^{0l})			,\							
															\tr( \ep_{jkl}\Ftil^{0j}D^k\Diso_{(n-1)}\Ftil^{0l})				$	&	$	d-3	$	\\	[4pt]
abelian	&	$	\cL_A^{(3)}			$	&		$	d=	3	$	&	$	\ep_{jkl} A^j F^{kl}				$	&		1		\\	[2pt]
	&	$	FF\cdots F			$	&		$	d=4n		$	&		any polynomial of order $n$ in $\von$, $\vtw$, $\vth$					&	$	\frac1{32}(d+4)(d+8)	$	\\	[2pt]
	&	$	F\prt_{(n-1)} F			$	&	even,	$	d\geq	4	$	&	$	F_{0j}\piso_{(n-1)}F^{0j}			,\							
															\Ftil_{0j}\piso_{(n-1)}\Ftil^{0j}				$	&	$	d-2	$	\\	[2pt]
	&	$	F\prt_{(n)} F			$	&	odd,	$	d\geq	5	$	&	$	\ep_{jkl}A^j \piso_{(n+1)}F^{kl}				$	&	$	\half (d-1)	$	\\	[2pt]
\hline
\hline
\end{tabular}
\end{table*}

Field redefinitions may also reduce the number
of independent Lorentz-invariant terms.
For example,
in the abelian limit of the spinor sector,
the quadratic term of the schematic form
$\psi \Ga_I D_{(2n-1)} \psi+\hc$
and $\psi \Ga_I D_{(2n-2)} \psi+\hc$
can be reduced via field redefinitions
to include only coefficients of the 
$a$, $c$, $g$, and $H$ types
\cite{km}.
Table \ref{tab:LI} then reveals
that the only remaining Lorentz-invariant term is
$\psb\ga^\mu D_\mu D^{\textrm{LI}}_{(2n-2)} \ps$,
which has even mass dimension $d$,
in agreement with known results
\cite{km}.

The literature contains several classical models 
of nonlinear electrodynamics that maintain Lorentz symmetry.
These models must therefore be special cases 
of the Lorentz-invariant abelian terms listed in Table \ref{tab:LI}.
One example is the Euler-Heisenberg lagrangian 
\cite{he36},
which is the effective one-loop Lagrange density
emerging from radiative corrections in QED.
The general form is a complicated function of $\PP$ and $\QQ$,
but in the weak-field limit it can be written in the form
\beq
\cL^{\textrm{EH}}=-\tfrac14 \PP+\frac{\al^2}{90m^4}(\PP^2+\tfrac74 \QQ^2),
\label{ehl}
\eeq
where $\al\simeq 1/137$ is the fine-structure constant
and $m$ is the mass of the electron.
This is a linear combination of Lorentz-invariant components
of the nonlinear pure-gauge terms with $d\leq 8$ 
shown in Table \ref{tab:LI}.
Note the absence of the parity-odd monomials $\QQ$ and $\PP \QQ$,
as required by the parity invariance of QED.

Another example is the Born-Infeld model 
\cite{bi34},
which introduces an upper bound for the electric field near the electron
that can eliminate the divergence of the self-energy.
In terms of $\PP$ and $\QQ$,
the Lagrange density of this model can be written as
\beq
\cL^{\textrm{BI}}=b^2-b^2\sqrt{1+\frac{\PP}{2b^2}-\frac{\QQ^2}{16b^4}},
\eeq
where $b$ is a scale parameter
representing the maximum attainable value of a pure electric field.
Expanding the square root for small $\PP$ and $\QQ$
reveals that the Lagrange density in this model 
can be viewed as a linear combination
of all the Lorentz-invariant components of the nonlinear pure-gauge terms
shown in Table \ref{tab:LI}
except those containing an odd power of $\QQ$.
The latter are parity odd,
and hence their omission again reflects the parity invariance
of the electromagnetic interaction.

\subsubsection{Isotropic limit}
\label{isotropic}

Models exhibiting isotropic physics can be generated 
by restricting the operators in the general nonabelian gauge theory
to those that preserve rotation invariance. 
Isotropic models are often adopted in the literature
due to their comparative simplicity.
Note, 
however,
that the isotropy can hold only in a specified inertial frame
and must be violated in other inertial frames.
Indeed,
Lorentz violation always implies rotation violation 
because boost transformations close into rotations under commutation.
Moreover,
the rotation of the Earth about its axis 
and the orbital revolution around the Sun 
imply that no laboratory is inertial,
so the set of isotropic operators is insufficient 
to characterize all physical effects even in isotropic models.
Nonetheless,
the isotropic limit is useful
in describing the subset of rotation-invariant effects.

Table \ref{tab:isotropic} lists the isotropic components
of some operators in gauge field theories.
The first column distinguishes the nonabelian and abelian scenarios.
The abelian fermion sector is obtained directly by restricting
the nonabelian one and so is omitted from the table.
Schematic forms for the operators are presented in the second column,
using $\Ga^I$ to denote a generic gamma-matrix structure
and suppressing all spacetime indices. 
The operator mass dimensions are listed in the third column,
with some values of $d$ specified in terms of an integer $n=1,2,\ldots$.
The fourth column displays the isotropic components
of the Lagrange density.
The last column provides the number of independent operators
except for the nonabelian combination $\tr(FF\cdots F)$,
which is complicated by the noncommutativity 
of products of the field strength.
Note that isotropic operators with an odd number of spacetime indices
are odd under CPT.

In the table,
all covariant derivatives are assumed to be totally symmetrized,
and we define
\beq
\von=F_{0j}F^{0j},
\quad
\vtw=F_{jk}F^{jk},
\quad
\vth=F_{0j}\Ftil^{0j}.
\eeq
By convention,
a ring diacritic on a quantity denotes its isotropic component.
Also,
the covariant derivative $\Dison$ represents a symmetrized monomial
of $n$ covariant derivatives involving products of $D_0$ and $D_j D^j$,
with a similar definition holding for the partial derivative $\pison$.
Both $\Dison$ and $\pison$ therefore contain 
$(n+2)/2$ independent isotropic operators for even $n$
and $(n+1)/2$ ones for odd $n$.

For the restriction to QED,
the field strength $F^{\mu\nu}$ is determined 
by the electric field $\mathbf{E}$ 
and the magnetic induction $\mathbf{B}$,
both of which have mass dimension two 
and transform under rotations as ordinary vectors.
All operators of the form $\tr(FF\cdots F)$ 
must therefore be polynomials formed from the three rotation invariants
$\von = -2 \mathbf{E}^2$,
$\vtw = 2 \mathbf{B}^2$,
$\vth = -2 \mathbf{E}\cdot\mathbf{B}$
and hence can exist only for $d=4n$.
For each $n$,
$(n+1)(n+2)/2$ independent operators exist.
In the full nonabelian theory,
more independent isotropic operators can be expected
because the three invariants have nontrivial commutators. 

Table \ref{tab:isotropic} also displays terms in the pure-gauge sector
of the schematic form $\tr(F\Dnn F)$,
which represent contributions to the gauge propagator.
The isotropic components of these terms 
can be obtained by adopting
$F^{0j}$ and $\Ftil^{0j}=\ep^{0jkl}F_{kl}/2$ 
as the fundamental variables contained in $F^{\mu\nu}$
and contracting indices to form rotation invariants. 
Some operators generated in this way,
such as $F^{0j}D_j D_k \Diso_{(n-2)} F^{0k}$ 
or $\ep_{jkl}F^{0j}D^k\Diso_{(n-1)}\Ftil^{0l}$,
are equivalent to others modulo surface terms 
when the homogeneous equation is imposed.
In the abelian limit,
the homogeneous equation further reduces the number of independent operators,
producing the results shown.
As before, 
the total number of independent operators 
may be further reduced via field redefinitions.

\section{Experiments}
\label{Experiments}

The previous section presents the construction 
of the general Lagrange density
for a nonabelian gauge field theory
describing Lorentz and CPT violation.
The construction contains
numerous predictions for physical effects.
In particular,
the special limiting cases of Lorentz-violating QED
and of Lorentz-violating QCD and QED coupled to quarks,
given explicitly in Tables \ref{tab:QED} and \ref{tab:QCD}, 
offer many prospects for experimental study.
To illustrate some of the possibilities,
we consider here two applications to experiments.
The first involves the effects on photon-photon scattering
of certain $d=8$ operators in Lorentz-violating QED,
and the second involves implications for deep inelastic scattering 
of some $d=5$ Lorentz- and CPT-violating operators 
in the QCD and QED interactions of quarks.

\subsection{Light-by-light scattering}
\label{Lightbylight}

Light-by-light scattering in the vacuum 
is a nonlinear process forbidden 
in classical linear Maxwell electrodynamics 
but predicted to occur in QED 
via radiative quantum corrections
and described by the Euler-Heisenberg Lagrange density \rf{ehl}
\cite{he36}.
Although the direct cross section for this process is tiny,
various techniques can be used to study it indirectly
\cite{photon2017}.
Direct observation of light-by-light scattering
has recently been demonstrated by the ATLAS collaboration
using ultraperipheral collisions of heavy ions in the LHC
\cite{ma17}.

The nonminimal sector of Lorentz-violating QED
given in Table \ref{tab:QED}
contains numerous operators that contribute
to the cross section for light-by-light scattering.
One-loop corrections due to some $d=5$ operators in the fermion sector
have been considered in Ref.\ \cite{gg18}. 
Here,
we study instead tree-level corrections to the cross section,
arising from $d=8$ photon-interaction operators in the pure-photon sector.
We calculate the effects of these corrections
on the experiment at the LHC
and use published data to obtain first constraints
on the corresponding coefficients for Lorentz violation.

\subsubsection{Setup}

The dominant SM contributions to light-by-light scattering
arise from one-loop radiative corrections.
In contrast,
the pure-photon sector of Lorentz-violating QED
contains nonlinear operators with four powers of the photon field
that yield tree-level contributions to light-by-light scattering
at leading order in Lorentz violation.
The dominant contribution of this type 
is expected to arise from the operator with mass dimension $d=8$. 
The corresponding contribution to the Lagrange density
is the last term in Eq.\ \rf{deight}.

\begin{figure}[htp]
\centering
\includegraphics[width=0.25\textwidth]{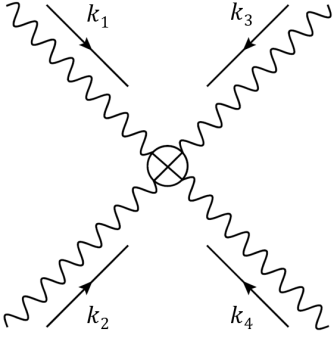}
\caption{Dominant contribution to light-by-light scattering}
\label{fig:Feynman}
\end{figure}

The Feynman diagram for the tree-level process in Lorentz-violating QED
is displayed in Fig.\ \ref{fig:Feynman}.
The Feynman rule 
\beq
-8 i (k_1)_{\mu_1}(k_2)_{\mu_2}(k_3)_{\mu_3}(k_4)_{\mu_4}
k_F^{(8)\mu_1\al_1\mu_2\al_2\mu_3\al_3\mu_4\al_4}
\eeq
depends on the momenta of the four photon lines
and is governed by the coefficient
$k_F^{(8)\mu_1\al_1\mu_2\al_2\mu_3\al_3\mu_4\al_4}$.
This coefficient is antisymmetric
under the exchange of indices $\mu_j \leftrightarrow \al_j$
for each $j=1,2,3,4$. 
It is also totally symmetric under exchanges of
any pair of indices $\mu_j\al_j$.
These symmetries imply that the coefficient
contains a total of 126 independent components,
each describing a distinct physical effect. 

The total cross section for light-by-light scattering is the integral 
of the spin-averaged squared modulus of the scattering amplitude 
over the phase space,
as usual.
In the laboratory frame,
we denote the four-momenta of the two incoming photons 
by $k_1$ and $k_2$
and the four-momenta of the two outgoing photons
by $k_3$ and $k_4$.
The spin-averaged squared modulus of the scattering amplitude
then takes the form 
\bea
&&
\hskip -5pt
\quar \Si \lvert \mathcal{T}_{fi} \rvert^2
= 16 \et_{\al_1\al_1\pr}\et_{\al_2\al_2\pr}
\et_{\al_3\al_3\pr}\et_{\al_4\al_4\pr}
\nn\\
&&
\hskip 10pt
\times 
(k_1)_{\mu_1}(k_1)_{\mu_1\pr}
(k_2)_{\mu_2}(k_2)_{\mu_2\pr}
(k_3)_{\mu_3}(k_3)_{\mu_3\pr}
(k_4)_{\mu_4}(k_4)_{\mu_4\pr}
\nn\\
&&
\hskip 10pt
\times 
k_F^{(8)\mu_1\al_1\mu_2\al_2\mu_3\al_3\mu_4\al_4}
k_F^{(8)\mu_1\pr\al_1\pr\mu_2\pr\al_2\pr\mu_3\pr\al_3\pr\mu_4\pr\al_4\pr}.
\label{amplitude}
\eea
Note that this expression is quadratic in the coefficient
$k_F^{(8)\mu_1\al_1\mu_2\al_2\mu_3\al_3\mu_4\al_4}$.

The result \rf{amplitude} exhibits an interesting duality symmetry.
Direct calculation reveals that it is invariant
under the replacement of
$k_F^{(8)\mu_1\al_1\mu_2\al_2\mu_3\al_3\mu_4\al_4}$
with its quadruple dual
$\widetilde{k}_F^{(8)\mu_1\al_1\mu_2\al_2\mu_3\al_3\mu_4\al_4}$
defined by
\bea
&&
\hskip-20pt
\widetilde{k}_F^{(8)\mu_1\al_1\mu_2\al_2\mu_3\al_3\mu_4\al_4} 
=\tfrac{1}{16} 
k^{(8)}_F{}_{\mu_1\pr\al_1\pr\mu_2\pr\al_2\pr\mu_3\pr\al_3\pr\mu_4\pr\al_4\pr}
\nn\\
&&
\hskip20pt
\times
\ep^{\mu_1\al_1\mu_1\pr\al_1\pr}
\ep^{\mu_2\al_2\mu_2\pr\al_2\pr}
\ep^{\mu_3\al_3\mu_3\pr\al_3\pr}
\ep^{\mu_4\al_4\mu_4\pr\al_4\pr}.
\label{quadual}
\eea
This symmetry guarantees that the experimental constraints 
on a component of the coefficient 
and the corresponding component of its dual are the same.

Denoting by $\Om$
the solid angle subtended by the momentum $\bf{k_3}$ relative to $\bf{k_1}$,
the theoretical differential cross section $d\si/d\Om$
can be written in the form
\beq
\frac {d\si} {d\Om} =
\frac{1}{256\pi^2 [(\om_1+\om_2)-(\om_1-\om_2)\cos \th]^2}
\Si \lvert \mathcal{T}_{fi} \rvert^2.
\label{gaga}
\eeq
In this expression,
the energies of the two incoming photons are $\om_1$ and $\om_2$,
and $\th$ is the angle between the momenta $\bf{k_1}$ and $\bf{k_3}$.
Note that the differential cross section
has nontrivial dependence on the azimuthal scattering angle $\ph$
arising from the Lorentz violation 
in the scattering amplitude $\mathcal{T}_{fi}$.

\subsubsection{Cross section and constraints}
\label{sec:sun}

\begin{figure}[htp]
\centering
\includegraphics[width=0.3\textwidth]{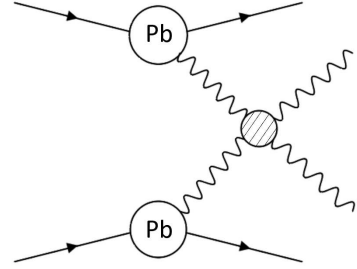}
\caption{Light-by-light scattering in ultraperipheral collisions}
\label{fig:lls}
\end{figure}

Our present interest is the application of the above derivation
to the data for light-by-light scattering
obtained from ultraperipheral collisions of lead ions
\cite{ma17}.
The relevant process is shown schematically in Fig.\ \ref{fig:lls},
which displays the role of the four-point photon vertex.
The photons from the high-energy lead ions 
can be viewed as photon beams in the equivalent photon approximation
\cite{epa}.

In the experiment,
the two incoming photons emitted by the lead ions are almost collinear.
The experimental cross section 
can be taken as a convolution of the differential cross section \rf{gaga} 
with the incoming photon-flux factors,
integrated over the range of observed solid angle
and outgoing photon energies
\cite{es16}.
It can be written in the form
\beq
\si^{\textrm{expt}}
=\int d\Om dm_{\ga\ga} dy_{\ga\ga} 
\frac {m_{\ga\ga}}{2}
\frac{n(\om_1)}{\om_1} \frac{n(\om_2)}{\om_2}
\frac{d\si}{d\Om},
\label{xptxsec}
\eeq
where the diphoton mass $m_{\ga\ga}=2\sqrt{\om_3\om_4}$
and the diphoton rapidity $y_{\ga\ga}=\ln(\om_4/\om_3)/2$
are functions of the energies $\om_3$ and $\om_4$ 
of the outgoing photons.
The photon-flux factors $n(\om_1)$ and $n(\om_2)$
depend on the incoming photon energies $\om_1$, $\om_2$,
which can be expressed in terms of $\om_3$, $\om_4$
and the scattering angle $\th$ by 
\bea
\om_1 &=&
\half \om_3 (1+\cos\th) 
+ \half \om_4 \left(1 \pm \sqrt{1-\frac{\om_3^2}{\om_4^2}\sin^2\th}\right),
\nn\\
\om_2 &=&
\half \om_3 (1-\cos\th) 
+ \half \om_4 \left(1 \mp \sqrt{1-\frac{\om_3^2}{\om_4^2}\sin^2\th}\right).
\qquad
\eea
For the calculation,
we assume $\om_3\leq\om_4$ for convenience.

\renewcommand\arraystretch{1.4}
\begin{table*}
\caption{
\label{tab:lls_iso}
Constraints from light-by-light scattering
on the modulus of combinations of photon coefficients.}
\setlength{\tabcolsep}{8pt}
\begin{tabular}{lll}
\hline
\hline
Limiting case &Coefficient combination & Constraint\\[2pt]
\hline
Lorentz invariant	&	$	\tfrac1{144}				\et_{\ka\mu}\et_{\la\nu}\et_{\rh\ta}\et_{\si\up}	k_F^{(8)	\ka\la\mu\nu\rh\si\ta\up	}	$	&	$	<	9.5	\times	10^{	-8	}	\textrm{ GeV}^{-4}	$	\\	[2pt]
	&	$	\tfrac1{576}				\ep_{\ka\la\mu\nu}\ep_{\rh\ta\si\up}	k_F^{(8)	\ka\la\mu\nu\rh\si\ta\up	}	$	&	$	<	9.5	\times	10^{	-8	}	\textrm{ GeV}^{-4}	$	\\	[2pt]
	&	$	\tfrac1{288}				\et_{\ka\mu}\et_{\la\nu}\ep_{\rh\ta\si\up}	k_F^{(8)	\ka\la\mu\nu\rh\si\ta\up	}	$	&	$	<	7.5	\times	10^{	-8	}	\textrm{ GeV}^{-4}	$	\\	[4pt]
Isotropic	&	$	\tfrac19	\sum_{	J,K	}		k_F^{(8)	TJTJTKTK	}	$	&	$	<	1.2	\times	10^{	-7	}	\textrm{ GeV}^{-4}	$	\\	[2pt]
	&	$	\tfrac19	\sum_{	J<K,L<M	}		k_F^{(8)	JKJKLMLM	}	$	&	$	<	1.2	\times	10^{	-7	}	\textrm{ GeV}^{-4}	$	\\	[2pt]
	&	$	\tfrac19	\sum_{	J,M,K<L,N<P	}	\ep_{JKL}\ep_{MNP}	k_F^{(8)	TJTMKLNP	}	$	&	$	<	9.5	\times	10^{	-8	}	\textrm{ GeV}^{-4}	$	\\	[2pt]
	&	$	\tfrac19	\sum_{	J,K<L	}		k_F^{(8)	TJTJKLKL	}	$	&	$	<	2.3	\times	10^{	-7	}	\textrm{ GeV}^{-4}	$	\\	[2pt]
	&	$	\tfrac19	\sum_{	J,K,L<M	}	\ep_{KLM}	k_F^{(8)	TJTJTKLM	}	$	&	$	<	5.6	\times	10^{	-8	}	\textrm{ GeV}^{-4}	$	\\	[2pt]
	&	$	\tfrac19	\sum_{	L,J<K,M<N	}	\ep_{LMN}	k_F^{(8)	JKJKTLMN	}	$	&	$	<	5.6	\times	10^{	-8	}	\textrm{ GeV}^{-4}	$	\\	[2pt]
\hline
\end{tabular}
\end{table*}

The ATLAS experiment 
\cite{ma17}
is sensitive to most of the $4\pi$ solid angle,
implying that the integral \rf{xptxsec} ranges over
$0.18\lsim\th\lsim 2.96$ and $0\leq\ph\leq2\pi$.
The data span the diphoton-mass range 
6 GeV$\lsim m_{\ga\ga} \lsim$ 24 GeV,
so we restrict the integral \rf{xptxsec} to this range
and adopt the corresponding diphoton-rapidity range
$0\lsim y_{\ga\ga} \lsim\ln(m_{\ga\ga}/6)$.
Unlike the SM result,
which is tiny for large diphoton masses, 
the maximum of $\si^{\textrm{expt}}$ in Lorentz-violating QED
is attained for a diphoton mass of around 600 GeV.
Our restriction on the upper value of the diphoton mass 
therefore ultimately translates into conservative constraints
on the coefficients $k_F^{(8)\mu_1\al_1\mu_2\al_2\mu_3\al_3\mu_4\al_4}$.
Future experiments at higher energies
can be expected to increase significantly
the sensitivity to these effects.

The photon-flux function $n(\om)$ 
is determined by the elastic form factor $F(q^2)$,
which is the Fourier transformation 
of the charge distribution of the nucleus.
It can be written as
\cite{gb02}
\beq
n(\om)=\frac{Z^2\al}{\pi^2}\int d^2q_\perp
\frac{q_\perp^2}{\big[(\om/\ga)^2+q_\perp^2\big]^2}
F^2\left(\left[\frac{\om}{\ga}\right]^2+q_\perp^2\right),
\label{fluxform}
\eeq
where $Z$ is the atomic number of the ion,
$\al\simeq /137$ is the fine-structure constant,
$\ga=1/\sqrt{1-\be^2}$ is the Lorentz factor of the nucleus,
$q_\perp$ is the transverse momentum,
and the integral of $q_\perp$ ranges over the whole plane.
Taking the form factor as
\cite{gb02}
$F(q^2)=\Th(1/R^2-q^2)$,
where $R$ is the nuclear radius,
yields
$n(\om)=({2Z^2\al}/{\pi})\ln({\ga}/{\om R})$
at leading order.
However,
this expression for the photon flux becomes negative for $\om>\ga/R$,
which is potentially problematic.
Instead,
we approximate $F(q^2)$ as a monopole form factor
\cite{htb94}
$F(q^2)={\La^2}/({\La^2+q^2})$,
where $\La=\sqrt{6/\langle r^2\rangle}$ 
and $\langle r^2\rangle$ 
is mean squared radius of the nucleus.
For ${}^{208}\textrm{Pb}$, 
experimental data 
\cite{vjv87} 
give $\langle r^2 \rangle^{1/2}=5.5$ fm and $\La=0.088$ GeV.
We find that the photon-flux function \rf{fluxform} then becomes
\beq
n(\om)=\frac{2Z^2\al}{\pi}
\left[ \frac{2(\om/\ga)^2+\La^2}{2\La^2} 
\ln\Big(1+\frac{\La^2}{(\om/\ga)^2}\Big)-1 \right].
\label{nomnew}
\eeq
This expression is positive for any $\om$
and converges to zero as $\om$ tends to infinity,
matching physical expectations.
For definiteness and simplicity,
we use the above expression to calculate the cross section.
As a cross check,
we have also used the fivefold integral presented in the literature
\cite{ks10} 
to estimate the photon flux and calculate the cross section,
obtaining results close to those using the expression \rf{nomnew}.

The LHC experiment measured the cross section as 
70$\pm$24(stat.)$\pm$17(syst.)~nb 
\cite{ma17}.
The theoretical SM prediction is 
49$\pm$10~nb 
\cite{ks10}.
The difference between these results is 21$\pm$31~nb,
which is compatible with zero.
An upper limit for the additional tree-level contribution
in Lorentz-violating QED
can therefore be taken as 83 nb at the 95\% confidence level.
We can use this result to constrain the coefficients
$k_F^{(8)\mu_1\al_1\mu_2\al_2\mu_3\al_3\mu_4\al_4}$.

The rotation of the Earth and its revolution about the Sun
imply that the laboratory is a noninertial frame,
which induces time dependence in the coefficients for Lorentz violation.
Experimental results on the coefficients
must therefore be reported in a prescribed inertial frame.
By standard convention in the literature,
this frame is taken to be the Sun-centered frame
\cite{sunframe}
with spatial coordinates $X^J$, $J=X,Y,Z$
chosen so that the $Z$ axis is aligned along the rotation axis of the Earth.
The $X$ axis points from the Earth to the Sun at the 2000 vernal equinox,
which is adopted as the origin of the time $T$.
The $Y$ axis completes the right-handed orthogonal coordinate system.
The contribution to the cross section therefore can be viewed
as determined by a nonlinear combination 
of the 126 independent components of the coefficient 
$k_F^{(8)\mu_1\al_1\mu_2\al_2\mu_3\al_3\mu_4\al_4}$
expressed in the Sun-centered frame.
The experimental result from the LHC
places a single constraint on this combination. 
However,
the result is unwieldy.
To gain insight,
we follow here the standard practice in the literature
of extracting constraints in various limiting scenarios.

\renewcommand\arraystretch{1.6}
\begin{table*}
\caption{
\label{tab:lls_one}
Constraints from light-by-light scattering
on the modulus of individual photon coefficients.}
\setlength{\tabcolsep}{5pt}
\begin{tabular}{ll}
\hline
\hline
Coefficient & Constraint\\
\hline
$	k_F^{(8)	TXTXTXTX	}, 	k_F^{(8)	TYTYTYTY	}, 	k_F^{(8)	XZXZXZXZ	}, 	k_F^{(8)	YZYZYZYZ	}	$	&	$	<	1.0	\times	10^{	-6	}	\textrm{ GeV}^{-4}	$	\\
$	k_F^{(8)	TZTZTZTZ	}, 	k_F^{(8)	XYXYXYXY							}	$	&	$	<	5.9	\times	10^{	-7	}	\textrm{ GeV}^{-4}	$	\\
$	k_F^{(8)	TXTXTXTY	}, 	k_F^{(8)	TYTYTYTX	}, 	k_F^{(8)	XZXZXZYZ	}, 	k_F^{(8)	YZYZYZXZ	}	$	&	$	<	6.4	\times	10^{	-7	}	\textrm{ GeV}^{-4}	$	\\
$	k_F^{(8)	TXTXTXTZ	}, 	k_F^{(8)	TYTYTYTZ	}, 	k_F^{(8)	XZXZXZXY	}, 	k_F^{(8)	YZYZYZXY	}	$	&	$	<	4.9	\times	10^{	-7	}	\textrm{ GeV}^{-4}	$	\\
$	k_F^{(8)	TXTXTXXY	}, 	k_F^{(8)	TYTYTYXY	}, 	k_F^{(8)	XZXZXZTZ	}, 	k_F^{(8)	YZYZYZTZ	}	$	&	$	<	5.5	\times	10^{	-7	}	\textrm{ GeV}^{-4}	$	\\
$	k_F^{(8)	TXTXTXXZ	}, 	k_F^{(8)	TYTYTYYZ	}, 	k_F^{(8)	XZXZXZTX	}, 	k_F^{(8)	YZYZYZTY	}	$	&	$	<	1.0	\times	10^{	-6	}	\textrm{ GeV}^{-4}	$	\\
$	k_F^{(8)	TXTXTXYZ	}, 	k_F^{(8)	TYTYTYXZ	}, 	k_F^{(8)	XZXZXZTY	}, 	k_F^{(8)	YZYZYZTX	}	$	&	$	<	5.2	\times	10^{	-7	}	\textrm{ GeV}^{-4}	$	\\
$	k_F^{(8)	TZTZTZTX	}, 	k_F^{(8)	TZTZTZTY	}, 	k_F^{(8)	XYXYXYXZ	}, 	k_F^{(8)	XYXYXYYZ	}	$	&	$	<	3.8	\times	10^{	-7	}	\textrm{ GeV}^{-4}	$	\\
$	k_F^{(8)	TZTZTZXY	}, 	k_F^{(8)	XYXYXYTZ							}	$	&	$	<	3.0	\times	10^{	-7	}	\textrm{ GeV}^{-4}	$	\\
$	k_F^{(8)	TZTZTZXZ	}, 	k_F^{(8)	TZTZTZYZ	}, 	k_F^{(8)	XYXYXYTX	}, 	k_F^{(8)	XYXYXYTY	}	$	&	$	<	4.2	\times	10^{	-7	}	\textrm{ GeV}^{-4}	$	\\
$	k_F^{(8)	TXTXTYTY	}, 	k_F^{(8)	XZXZYZYZ							}	$	&	$	<	5.6	\times	10^{	-7	}	\textrm{ GeV}^{-4}	$	\\
$	k_F^{(8)	TXTXTZTZ	}, 	k_F^{(8)	TYTYTZTZ	}, 	k_F^{(8)	XYXYXZXZ	}, 	k_F^{(8)	XYXYYZYZ	}	$	&	$	<	3.6	\times	10^{	-7	}	\textrm{ GeV}^{-4}	$	\\
$	k_F^{(8)	TXTXXYXY	}, 	k_F^{(8)	TYTYXYXY	}, 	k_F^{(8)	TZTZXZXZ	}, 	k_F^{(8)	TZTZYZYZ	}	$	&	$	<	9.7	\times	10^{	-7	}	\textrm{ GeV}^{-4}	$	\\
$	k_F^{(8)	TXTXXZXZ	}, 	k_F^{(8)	TYTYYZYZ							}	$	&	$	<	1.1	\times	10^{	-6	}	\textrm{ GeV}^{-4}	$	\\
$	k_F^{(8)	TXTXYZYZ	}, 	k_F^{(8)	TYTYXZXZ							}	$	&	$	<	4.3	\times	10^{	-7	}	\textrm{ GeV}^{-4}	$	\\
$	k_F^{(8)	TZTZXYXY										}	$	&	$	<	2.4	\times	10^{	-7	}	\textrm{ GeV}^{-4}	$	\\
$	k_F^{(8)	TXTXTYTZ	}, 	k_F^{(8)	TYTYTXTZ	}, 	k_F^{(8)	XZXZXYYZ	}, 	k_F^{(8)	YZYZXYXZ	}	$	&	$	<	3.4	\times	10^{	-7	}	\textrm{ GeV}^{-4}	$	\\
$	k_F^{(8)	TXTXTYXY	}, 	k_F^{(8)	TYTYTXXY	}, 	k_F^{(8)	XZXZTZYZ	}, 	k_F^{(8)	YZYZTZXZ	}	$	&	$	<	4.0	\times	10^{	-7	}	\textrm{ GeV}^{-4}	$	\\
$	k_F^{(8)	TXTXTYXZ	}, 	k_F^{(8)	TYTYTXYZ	}, 	k_F^{(8)	XZXZTXYZ	}, 	k_F^{(8)	YZYZTYXZ	}	$	&	$	<	5.2	\times	10^{	-7	}	\textrm{ GeV}^{-4}	$	\\
$	k_F^{(8)	TXTXTYYZ	}, 	k_F^{(8)	TYTYTXXZ	}, 	k_F^{(8)	XZXZTYYZ	}, 	k_F^{(8)	YZYZTXXZ	}	$	&	$	<	4.2	\times	10^{	-7	}	\textrm{ GeV}^{-4}	$	\\
$	k_F^{(8)	TXTXTZXY	}, 	k_F^{(8)	TYTYTZXY	}, 	k_F^{(8)	YZYZTZXY	}, 	k_F^{(8)	XZXZTZXY	}	$	&	$	<	2.8	\times	10^{	-7	}	\textrm{ GeV}^{-4}	$	\\
$	k_F^{(8)	TZTZTXYZ	}, 	k_F^{(8)	TZTZTYXZ	}, 	k_F^{(8)	XYXYTXYZ	}, 	k_F^{(8)	XYXYTYXZ	}	$	&	$	<	2.8	\times	10^{	-7	}	\textrm{ GeV}^{-4}	$	\\
$	k_F^{(8)	TXTXTZXZ	}, 	k_F^{(8)	TYTYTZYZ	}, 	k_F^{(8)	XZXZTXXY	}, 	k_F^{(8)	YZYZTYXY	}	$	&	$	<	5.1	\times	10^{	-7	}	\textrm{ GeV}^{-4}	$	\\
$	k_F^{(8)	TXTXTZYZ	}, 	k_F^{(8)	TYTYTZXZ	}, 	k_F^{(8)	XZXZTYXY	}, 	k_F^{(8)	YZYZTXXY	}	$	&	$	<	2.9	\times	10^{	-7	}	\textrm{ GeV}^{-4}	$	\\
$	k_F^{(8)	TXTXXYXZ	}, 	k_F^{(8)	TYTYXYYZ	}, 	k_F^{(8)	XZXZTXTZ	}, 	k_F^{(8)	YZYZTYTZ	}	$	&	$	<	6.0	\times	10^{	-7	}	\textrm{ GeV}^{-4}	$	\\
$	k_F^{(8)	TXTXXYYZ	}, 	k_F^{(8)	TYTYXYXZ	}, 	k_F^{(8)	XZXZTYTZ	}, 	k_F^{(8)	YZYZTXTZ	}	$	&	$	<	3.0	\times	10^{	-7	}	\textrm{ GeV}^{-4}	$	\\
$	k_F^{(8)	TXTXXZYZ	}, 	k_F^{(8)	TYTYXZYZ	}, 	k_F^{(8)	XZXZTXTY	}, 	k_F^{(8)	YZYZTXTY	}	$	&	$	<	4.8	\times	10^{	-7	}	\textrm{ GeV}^{-4}	$	\\
$	k_F^{(8)	TZTZTXTY	}, 	k_F^{(8)	XYXYXZYZ							}	$	&	$	<	2.9	\times	10^{	-7	}	\textrm{ GeV}^{-4}	$	\\
$	k_F^{(8)	TZTZTXXY	}, 	k_F^{(8)	TZTZTYXY	}, 	k_F^{(8)	XYXYTZXZ	}, 	k_F^{(8)	XYXYTZYZ	}	$	&	$	<	2.3	\times	10^{	-7	}	\textrm{ GeV}^{-4}	$	\\
$	k_F^{(8)	TZTZTXXZ	}, 	k_F^{(8)	TZTZTYYZ	}, 	k_F^{(8)	XYXYTXXZ	}, 	k_F^{(8)	XYXYTYYZ	}	$	&	$	<	3.9	\times	10^{	-7	}	\textrm{ GeV}^{-4}	$	\\
$	k_F^{(8)	TZTZXYXZ	}, 	k_F^{(8)	TZTZXYYZ	}, 	k_F^{(8)	XYXYTXTZ	}, 	k_F^{(8)	XYXYTYTZ	}	$	&	$	<	2.4	\times	10^{	-7	}	\textrm{ GeV}^{-4}	$	\\
$	k_F^{(8)	TZTZXZYZ	}, 	k_F^{(8)	XYXYTXTY							}	$	&	$	<	7.3	\times	10^{	-7	}	\textrm{ GeV}^{-4}	$	\\
$	k_F^{(8)	TXTYTZXY	}, 	k_F^{(8)	TZXYXZYZ							}	$	&	$	<	2.2	\times	10^{	-7	}	\textrm{ GeV}^{-4}	$	\\
$	k_F^{(8)	TXTYTZXZ	}, 	k_F^{(8)	TXTYTZYZ	}, 	k_F^{(8)	TXXYXZYZ	}, 	k_F^{(8)	TYXYXZYZ	}	$	&	$	<	2.9	\times	10^{	-7	}	\textrm{ GeV}^{-4}	$	\\
$	k_F^{(8)	TXTYXYXZ	}, 	k_F^{(8)	TXTYXYYZ	}, 	k_F^{(8)	TXTZXZYZ	}, 	k_F^{(8)	TYTZXZYZ	}	$	&	$	<	3.2	\times	10^{	-7	}	\textrm{ GeV}^{-4}	$	\\
$	k_F^{(8)	TXTYXZYZ										}	$	&	$	<	3.4	\times	10^{	-7	}	\textrm{ GeV}^{-4}	$	\\
$	k_F^{(8)	TXTZXYXZ	}, 	k_F^{(8)	TYTZXYYZ							}	$	&	$	<	3.8	\times	10^{	-7	}	\textrm{ GeV}^{-4}	$	\\
$	k_F^{(8)	TXTZXYYZ	}, 	k_F^{(8)	TYTZXYXZ							}	$	&	$	<	3.5	\times	10^{	-7	}	\textrm{ GeV}^{-4}	$	\\
\hline
\end{tabular}
\end{table*}

Consider first the Lorentz-invariant limit.
Inspection of Table \ref{tab:LI} reveals that there are
three Lorentz-invariant combinations of components of 
the coefficient $k_F^{(8)\mu_1\al_1\mu_2\al_2\mu_3\al_3\mu_4\al_4}$.
The orientation and velocity of the laboratory 
relative to the canonical Sun-centered frame 
plays no role for these combinations,
so constraints can be derived directly from the experimental data.
The first three rows of Table \ref{tab:lls_iso} 
lists the results of this analysis.
Each row specifies a Lorentz-invariant coefficient combination
and the corresponding constraint,
obtained under the assumption 
that other independent coefficient combinations vanish.
The fractions appearing in front of the combinations
are chosen to insure the combination has the same weight
as a single component of $k_F^{(8)\mu_1\al_1\mu_2\al_2\mu_3\al_3\mu_4\al_4}$.
In terms of the invariants \rf{pqd},
the operators in the Lagrange density
associated with the first, second, and third row
are $\PP^2$, $\QQ^2$, and $\PP\QQ$,
respectively. 

We can also consider the isotropic limit of Lorentz-violating QED.
According to the results in Table \ref{tab:isotropic},
the coefficient $k_F^{(8)\mu_1\al_1\mu_2\al_2\mu_3\al_3\mu_4\al_4}$
contains six isotropic combinations of components.
Since isotropic effects are independent of the laboratory orientation
and since the laboratory boost $\be\simeq 10^{-4}$ is negligible
in the Sun-centered frame,
we can again obtain constraints on the isotropic combinations 
directly from an analysis performed in the laboratory.
The resulting constraints in the Sun-centered frame
are displayed in the second part of Table \ref{tab:lls_iso}.
The isotropic combinations are listed in the second column,
where the summations are over spatial indices $J$, $K$, $\ldots = X,Y,Z$.
The third column presents the corresponding constraint
on the modulus of each combination
expressed in the Sun-centered frame,
determined assuming other coefficient components vanish.
The constraints on the first two and last two isotropic combinations 
are identical,
as a consequence of the duality symmetry 
associated with the quadruply dualized coefficients \rf{quadual}.

To implement a complete analysis,
the cross section expressed using coefficients in the laboratory frame
must be transformed to the Sun-centered frame.
To a good approximation,
the small Earth boost can be neglected in the transformation. 
Denoting the spatial coordinates in the laboratory frame by $x^j$,
the rotation matrix between the two frames is given by 
\beq
R^{jJ}= \left(
\begin{array}{ccc}
\cos\ch \cos\om_\oplus T_\oplus & \cos\ch \sin\om_\oplus T_\oplus & -\sin\ch
\\
-\sin\om_\oplus T_\oplus & \cos \om_\oplus T_\oplus & 0 
\\
\sin \ch \cos \om_\oplus T_\oplus & \sin \ch \sin\om_\oplus T_\oplus 
& \cos \ch
\end{array} \right),
\label{rotz}
\eeq
where $\om_\oplus\simeq 2\pi/(23{\rm ~h} ~56{\rm ~min})$
is the Earth's sidereal rotation frequency,
$T_\oplus$ is a convenient local sidereal time,
and $\ch$ is the angle between the beam direction and the $Z$ axis.
For the ATLAS detector at the LHC,
$\ch\simeq 81.4^\circ$.

Implementing this transformation on the cross section
generates a lengthy expression
containing up to fourth harmonics of the sidereal frequency.
The measured cross section is therefore predicted
to display sidereal variations in time at all these harmonics.
Analyzing experimental data to search for these time variations
would be of great interest,
but insufficient data are presently available to implement this.
Instead,
we consider here the time-averaged cross section,
which selects the time-independent contributions for consideration.
Calculation reveals that all 126 independent components of 
$k_F^{(8)\mu_1\al_1\mu_2\al_2\mu_3\al_3\mu_4\al_4}$
contribute to this result. 

Table \ref{tab:lls_one} presents the constraints on all 126 components,
taken one at a time with all others set to zero.
The first column lists the components,
and the second column shows the constraint
on the modulus of each one.
These results are the first in the literature
on nonlinear Lorentz-violating photon interactions.
Existing constraints on $d=8$ pure-photon terms in Lorentz-violating QED
are restricted to effects on photon propagation
in the astrophysical and laboratory contexts
\cite{km,d8astro,d8lab}.
Inspection confirms that the constraints in the table
exhibit the duality symmetry associated with Eq.\ \rf{quadual},
as expected.
The results also reveal symmetry under interchange of the $X$ and $Y$ indices.
This is a consequence of the rotation of the Earth about the $Z$ axis,
which implies that the time-averaged cross section 
must be invariant under rotation in the $X$-$Y$ plane
and hence under the exchange $(X,Y)\leftrightarrow (Y,-X)$.
The sign in this exchange has no effect on the constraints
because the cross section depends only on the square of the coefficients.

\subsection{Deep inelastic scattering}
\label{Deep inelastic scattering}

Deep inelastic scattering  
offers crucial experimental support for the existence of quarks
and the predictions of QCD,
and it serves as a tool in searches for physics beyond the SM 
\cite{dis,disproc}.
An initial study of the effects of Lorentz violation
on deep inelastic electron-proton scattering
\cite{klv17}
has studied the use of DIS
to search for Lorentz-violating effects of $c$-type coefficients
in the QCD and QED interactions of quarks.
This work complements recent theoretical and experimental efforts
to investigate Lorentz and CPT violation in the quark sector 
involving operators of mass dimensions $d=3$
\cite{ak98,d3quark},
$d=4$
\cite{d4quark},
and $d=5$
\cite{nvt16}.
The construction of the nonminimal operators
given in Table \ref{tab:QCD}
offers numerous interesting prospects for further exploration
in Lorentz-violating QCD.

Here,
we illustrate the possibilities by investigating
the effects on the DIS cross section
of a subset of nonminimal Lorentz- and CPT-violating terms 
in the Lagrange density for QCD and QED coupled to quarks.
Following the general approach presented 
in Ref.\ \cite{klv17},
we focus on unpolarized electron-proton DIS,
for which it suffices to limit attention to spin-independent operators.
For definiteness,
we work with spin-independent operators
of lowest nonminimal mass dimension,
which turn out to be governed 
by the $a^{(5)}$- and $a_F^{(5)}$-type coefficients.
Calculation reveals that the $a_F^{(5)}$-type coefficients
leave unaffected the DIS cross section at leading order,
so in what follows we focus on the experimentally measurable signals
produced by the $a^{(5)}$-type coefficients.

\subsubsection{Setup}

For electron-proton DIS,
the relevant terms in the Lagrange density 
that involve the valence quarks in the proton are 
\bea
&&
\hskip-25pt
-\half \sum_{f} a_f^{(5)\mu\al\be} 
\psb_f \ga_\mu iD_{(\al} iD_{\be)}\ps_f+\hc 
\nn\\
&=& 
- \sum_{f} a_f^{(5)\mu\al\be}
\big[\psb_f \ga_\mu i\prt_\al i \prt_\be \ps_f
-q_fA_{\al} \psb_f \ga_\mu i\prt_{\be} \ps_f 
\nn\\
&&
\hskip 25pt
+q_fA_{\al} (i\prt_{\be} \psb_f)\ga_\mu \ps_f
+q_f^2 A_\al A_\be \psb_f \ga_\mu \ps_f \big],
\eea
where some surface terms are neglected.
In this expression,
we limit attention to flavor-diagonal effects
and the two dominant quark flavors $f=u$, $d$.
The coefficients $a^{(5)\mu\al\be}_f$ are symmetric 
under interchange of the last two indices,
$a^{(5)\mu\al\be}_f=a^{(5)\mu\be\al}_f$.

At leading order,
the DIS process of interest involves the exchange of a photon
between the scattered electron and the proton.
We can therefore in principle expect Lorentz-violating effects 
in the photon and electron sectors to play a role, 
along with potential effects caused by the binding of the valence quarks
to form the proton.
For simplicity and definiteness,
we neglect here effects from the photon and electron sectors,
many of which are already tightly bounded
\cite{tables}.
Note in particular that existing limits on $d=5$ operators 
in the photon sector lie far below DIS sensitivities
\cite{km,photon}.
Also,
while precision studies of $d=5$ operators in the electron sector
are less broadly constrained
\cite{km,ms16,kv18},
they nonetheless also lie below the DIS sensitivities estimated below.
In contrast,
in the $d=5$ proton sector the coefficients 
$a^{(5)\mu\al\be}_p$ 
are only partially bounded 
\cite{kv18},
and the existing DIS sensitivities are roughly comparable 
to those estimated below.
Since the structure of the proton is nontrivial 
and little is known about Lorentz and CPT violation 
in the strong binding by the gluons and sea species,
we treat the coefficients 
$a^{(5)\mu\al\be}_f$ and $a^{(5)\mu\al\be}_p$ 
as independent here.

Coefficients of the $a^{(5)}$ type affect various particle properties.
Consider first effects at the level of relativistic quantum mechanics.
In the presence of these Lorentz- and CPT-violating effects,
the Dirac equation for a freely propagating fermion is modified to
\beq
(i \ga_\mu \prt^\mu 
- a^{(5)\mu\al\be}\ga_\mu i\prt_{\al} i\prt_{\be}-m)\ps=0.
\label{modir}
\eeq
For perturbative Lorentz violation,
this equation has two positive-energy and two negative-energy solutions.
As usual,
we reinterpret the negative-energy solutions 
as positive-energy antiparticle solutions with reversed momentum.
The plane-wave eigenfunctions can be written as 
$\ps^{(s)}(x)=\exp(-ip_u\cdot x)u^{(s)}({\bf p},a)$
for the positive-energy solutions and 
$\ps^{(s)}(x)=\exp(+ip_v\cdot x)v^{(s)}({\bf p},a)$
for the negative-energy ones,
where $p_u^\mu=(E_u,\bf{p})$ and $p_v^\mu=(E_v,\bf{p})$ 
are the four-momenta of the particle and antiparticle,
$\bf{p}$ is the three momentum,
and $s=1,2$ labels the spin.

Since the Lorentz-violating terms in the Lagrange density contain
only spin-independent operators,
the energies $E_u$ and $E_v$ are independent of the spins.
However,
the Lorentz violation typically implies that $E_u\neq E_v$,
as can be seen from the dispersion relations
\bea
(p_u^\mu-a^{(5)\mu p_up_u})\et_{\mu\nu}(p_u^\nu-a^{(5)\nu p_up_u})&=&m^2,
\nn\\
(p_v^\mu+a^{(5)\mu p_vp_v})\et_{\mu\nu}(p_v^\nu+a^{(5)\nu p_vp_v})&=&m^2.
\label{dispersion}
\eea
In these expressions,
an index $p_u$ or $p_v$ 
implies contraction with the corresponding four-momentum.
For example,
we write $a^{(5)\mu p_up_u}\equiv a^{(5)\mu\al\be}p_{u\al} p_{u\be}$.
The dispersion relations \rf{dispersion} reveal
that the energies of the particle and antiparticle
are related by $E_u({\bf p},a)=E_v({\bf p},-a)$,
as is expected given that the $a^{(5)}$-type coefficients
govern CPT-odd operators.

The momentum-space solutions of the modified Dirac equation \rf{modir}
can be written as
\bea
u^{(s)}({\bf p},a)&=&\frac{p_u^\mu \ga_\mu-a^{(5)\mu p_u p_u} \ga_\mu+m}
{\sqrt{E_u-a^{(5)0 p_u p_u}+m}} u^{(s)}_0,
\nn\\
v^{(s)}({\bf p},a)&=&\frac{-p_v^\mu \ga_\mu-a^{(5)\mu p_v p_v} \ga_\mu+m}
{\sqrt{E_v+a^{(5)0 p_v p_v}+m}} v^{(s)}_0,
\label{uv}
\eea
where $u^{(1)}_0{}\da=(1,0,0,0)$, 
$u^{(2)}_0{}\da=(0,1,0,0)$ 
and similarly $v^{(1)}_0{}\da=(0,0,1,0)$, 
$v^{(2)}_0{}\da=(0,0,0,1)$.
These solutions can be obtained from the conventional Lorentz-invariant ones
by the substitutions 
$p^\mu\to p_u^\mu-a^{(5)\mu p_u p_u}$ 
and 
$p^\mu\to p_v^\mu+a^{(5)\mu p_v p_v}$.
They therefore obey the usual normalization relations,
\beq
\ov{u}^{(s)}({\bf p},a) u^{(s\pr)}({\bf p},a)
=-\ov{v}^{(s)}({\bf p},a)v^{(s\pr)}({\bf p},a)
=2m\de^{ss\pr}.
\eeq
However,
the traditional orthogonality relations are replaced by 
\beq
\overline{u}^{(s)}({\bf p},a) v^{(s\pr)}({\bf p},-a)
=\overline{v}^{(s)}({\bf p},a)u^{(s\pr)}({\bf p},-a)=0,
\eeq
incorporating a sign change of the coefficients.

The solutions \rf{uv} imply modifications to the conventional spin sums.
We find
\bea
\sum_s u^{(s)}({\bf p},a) \overline{u}^{(s)}({\bf p},a)&=&
p_u^{\al} \ga_\al-a^{(5)\al p_u p_u} \ga_\al+m,
\nn\\
\sum_s v^{(s)}({\bf p},a) \overline{v}^{(s)}({\bf p},a)&=&
p_v^{\al} \ga_\al+a^{(5)\al p_v p_v} \ga_\al-m.
\qquad
\label{spinsum}
\eea
As in the usual Lorentz-invariant case,
we can write two projection operators from these expressions, 
\bea
\La_u({\bf p},a)&=&
\frac{1}{2m}\sum_s u^{(s)}({\bf p},a) \overline{u}^{(s)}({\bf p},a),
\nn\\
\La_v({\bf p},a)&=&
-\frac{1}{2m}\sum_s v^{(s)}({\bf p},a) \overline{v}^{(s)}({\bf p},a).
\eea
As required,
both these operators are idempotent,
$\La_u({\bf p},a)^2=\La_u({\bf p},a)$
and $\La_v({\bf p},a)^2=\La_v({\bf p},a)$,
and hence can be used as projectors in the two energy subspaces.
However,
they are no longer directly complementary.
Instead,
we find $\La_u({\bf p},a)+\La_v({\bf p},-a)=1$.

\begin{figure}[htp]
\centering
\includegraphics[width=0.45\textwidth]{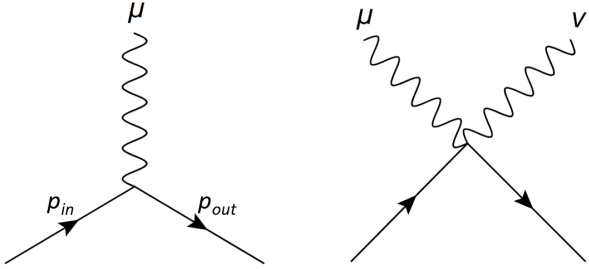}
\caption{The $A\ps\ps$ vertex (left) 
and the $AA\ps\ps$ vertex (right).}
\label{fig:vertex}
\end{figure}

At the level of the quantum field theory,
the modified Feynman propagator is found to be 
\beq
iS_{F}(p)=
\frac{i}{\ga_\mu p^\mu - a^{(5)\mu\al\be}\ga_\mu p_\al p_\be-m+ i\ep}.
\eeq
The Feynman rule for the photon-fermion-fermion vertex 
shown in Fig.\ \ref{fig:vertex} becomes
\beq
-iq\Ga^\mu(p_{\textrm{in}},p_{\textrm{out}})
=-iq[\ga^\mu - a^{(5)\al\be\mu}\ga_\al (p_{\textrm{in}}+p_{\textrm{out}})_\be],
\eeq
where $p_{\textrm{in}}$ and $p_{\textrm{out}}$ 
are the four-momenta of the incoming and outgoing fermions.
The coefficient $a^{(5)}{}^{\al\be\mu}$ 
is also associatd with a new vertex 
involving two photon lines and two fermion lines,
as displayed in Fig.\ \ref{fig:vertex}.
The corresponding Feynman rule is 
\beq
iq^2 \Ga^{(4)\mu\nu}=-2iq^2 a^{(5)\al\mu\nu} \ga_\al.
\label{fourlinevertex}
\eeq
This vertex does not contribute to the DIS process at tree level,
but it must be included if loops are considered.

\subsubsection{Cross section and constraints}

The DIS differential cross section
is a function of the phase-space variables
for the scattered electron.
Taking the $z$ axis as aligned with the incoming beam direction
and denoting by $\th$, $\ph$ the scattering angles of the electron 
in spherical polar coordinates,
the four-momenta for the incoming electron, 
incoming proton, and outgoing electron can be written as
$k^\mu=E(1,0,0,1)$, $p^\mu=E_p(1,0,0,-1)$,
and $k^{\prime\mu}=E\pr(1,\sin\th \cos\ph,\sin\th \sin\ph,\cos\th)$,
respectively.
The intermediate photon then has four-momentum $q=k-k\pr$.
It is convenient to introduce
the standard Mandelstam $s$ and Bjorken $x$ and $y$ variables via 
\beq
s=(p+k)^2,
\quad
x=\frac{-q^2}{2p\cdot q},
\quad
y=\frac{p\cdot q}{p\cdot k}.
\eeq
In the expressions to follow,
we can safely disregard the proton mass $M$
because in DIS experiments $M^2\ll Q^2=-q^2$ and $M^2\ll 2p\cdot q$.
The quark masses can similarly be neglected.

In the presence of Lorentz violation,
the unpolarized differential cross section 
for the DIS process can be written in the form
\cite{klv17}
\beq
\frac{d\si}{dxdyd\ph}=
\frac{\al^2 y}{2\pi q^4} L_{\mu\nu} \Im W^{\mu\nu},
\eeq
where $\al$ is the fine structure constant,
$L_{\mu\nu}=2(k_\mu k\pr_\nu+k_\nu k\pr_\mu - k\cdot k\pr \et_{\mu\nu})$ 
is the electron tensor,
and $W^{\mu\nu}$ is the proton tensor.
To calculate the explicit form of $W^{\mu\nu}$,
we use the parton model.
Since the coefficients $a^{(5)\mu\al\be}$
are assumed perturbatively small,
we keep only terms at first order in what follows.

In the standard parton model,
the parton momentum $p_f=\xi p$ is taken to be a fraction $\xi$
of the proton momentum $p$.
However,
the modified dispersion relation \rf{dispersion}
enforces the conditions 
$p^2=2a_p^{(5)}{}^{ppp}$ 
and 
$p_f^2=2a_f^{(5)}{}^{p_fp_fp_f}$,
which are incompatible with $p_f=\xi p$.
Instead,
we assume $p_f^\mu=\xi p^\mu+\de_f^\mu$
with a small momentum correction $\de_f^\mu$.
Noting that $p\cdot \de_f=\xi^2 a_f^{(5)}{}{ppp}-\xi a_p^{(5)}{}^{ppp}$,
we find
\beq
p_f^\mu=\xi p^\mu-\xi a_p^{(5)\mu pp}+\xi^2 a_f^{(5)\mu pp}.
\label{parton}
\eeq
The implication of this shift for the parton distribution functions
is an interesting topic for future investigation.

Following the treatment 
in Ref.\ \cite{klv17},
we can use the parton model and the optical theorem
to obtain the explicit form of $\Im W^{\mu\nu}$.
The contribution involving the $AA\ps\ps$ vertex 
shown in Fig.\ \ref{fig:vertex}
is purely real and so is irrelevant in this context.
After some calculation,
we find that the contribution involving the $A\ps\ps$ vertex is 
\bea
W^{\mu\nu}&=&-\half \int_0^1 d\xi \sum_f q_f^2 \frac{f_f(\xi)}{\xi} 
\nn\\
&&
\hskip -30pt
\times 
\tr \Big[\ga_\al (p_f^{\al}-a_f^{(5)\al p_f p_f}) 
\big(\ga^\mu-a_f^{(5)\al\pr\be\pr\mu}\ga_{\al\pr}(2p_f+q)_{\be\pr}\big)
\nn\\
&&
\times
\ga_\be \big(p_f^\be+q^\be-a_f^{(5)\be\rh\si} (p_f+q)_\rh (p_f+q)_\si \big)
\nn\\
&&
\times
\big(\ga^\nu-a_f^{(5)\al^{\prime\prime}\be^{\prime\prime}\nu}
\ga_{\al^{\prime\prime}}(2p_f+q)_{\be^{\prime\prime}}\big)
\Big]
\nn\\
&&
\hskip -30pt
\times
\Big(\big[ \ga_\be \big(p_f^\be+q^\be
-a_f^{(5)\be\rh\si} (p_f+q)_\rh (p_f+q)_\si \big)\big]^2
+i\ep\Big)^{-1} 
\nn\\
&&
+(\mu \leftrightarrow \nu, q\leftrightarrow -q),
\label{wmn}
\eea
We are interested in the imaginary part of this expression.
This comes from the propagator factor,
\bea
&&
\textrm{Im}\big(\big[\ga_\be 
\big(p_f^\be+q^\be-a_f^{(5)\be\rh\si} 
(p_f+q)_\rh (p_f+q)_\si \big)\big]^2+i\ep\big)^{-1}
\nn\\
&&
\hskip 20pt
=-\de_f \de(\xi-x\pr_f),
\eea
where
\bea
\de_f &=&
\frac{\pi}{ys}
\bigg[1+\frac{2}{s}a_p^{(5)}{}^{ppp}
\nn\\
&&
+\frac{2}{ys}
(a_p^{(5)}{}^{qpp}+4xa_f^{(5)}{}^{ppq}+2a_f^{(5)}{}^{qqp}+a_f^{(5)}{}^{pqq})
\bigg],
\nn\\
\eea
and $x\pr_f=x-x_f$ with
\bea
x_f&=&
-\frac{2}{ys}(x a_p^{(5)}{}^{qpp}+2x^2 a_f^{(5)}{}^{ppq}
\nn\\
&&
\hskip 30pt
+2x a_f^{(5)}{}^{qqp}+x a_f^{(5)}{}^{pqq}+a_f^{(5)}{}^{qqq}).
\eea

As a check on these calculations,
we have used the explicit results for $\de_f$ and $x_f$
to verify the Ward identity $q_\mu \textrm{Im} W^{\mu\nu}=0$.
This requires incorporating
both the modifications in the propagator and in the vertex,
which together insure the preservation of gauge invariance.
Note that the photon-photon-fermion-fermion vertex \rf{fourlinevertex}
can be omitted from the calculation
because it contributes only at loop level.
We also remark that the modification \rf{parton} of the parton momentum 
is crucial for the Ward identity to hold.

After contracting the result for the imaginary part 
of the proton tensor \rf{wmn}
with the electron tensor $L_{\mu\nu}$,
some calculation yields the differential cross section as
\bea
\frac{d\si}{dxdyd\ph}
&=&
\frac{\al^2}{q^4}\sum_f F_{2f} 
\bigg[ \frac{ys^2}{\pi}[1+(1-y)^2]\de_f
\nn\\
&&
\hskip 20pt
+\frac{y(y-2)s}{x}x_f
-4[1+(1-y)^2]a_p^{(5)}{}^{ppp}
\nn\\
&&
\hskip 20pt
+4(y-2)a_p^{(5)}{}^{kpp}-2y a_p^{(5)}{}^{qpp}
\nn\\
&&
\hskip 20pt
-\frac{4}{x}(4x^2 a_f^{(5)}{}^{ppk}+2x a_f^{(5)}{}^{pkq}+2x a_f^{(5)}{}^{qkp}
\nn\\
&&
\hskip 20pt
\hskip 20pt
+2x a_f^{(5)}{}^{kpq}+a_f^{(5)}{}^{qqk}+a_f^{(5)}{}^{kqq})
\nn\\
&&
\hskip 20pt
+2y(4x^2 a_f^{(5)}{}^{ppp}+2x a_f^{(5)}{}^{ppq}+2x a_f^{(5)}{}^{qpp}
\nn\\
&&
\hskip 20pt
\hskip 20pt
+a_f^{(5)}{}^{qqp}+4x a_f^{(5)}{}^{ppk}+2 a_f^{(5)}{}^{pkq})
\nn\\
&&
\hskip 20pt
-y^2 s~ \et_{\mu\nu} (2x a_f^{(5)\mu\nu p}+a_f^{(5)\mu\nu q})
\nn\\
&&
\hskip 20pt
+\frac{4y}{x}(2x a_f^{(5)}{}^{kkp}+a_f^{(5)}{}^{kkq})
\bigg],
\label{discross}
\eea
where $F_{2f}=q_f^2f_f(x\pr_f)x\pr_f$.
The structure of this result has enticing similarities
to the differential cross section 
obtained for $c$-type coefficients in Eq.\ (14)
of Ref.\ \cite{klv17}. 
The role of the coefficients $c_f^{pq}$ in that equation
parallels the role of the coefficients $a^{(5)}{}^{kpq}$ here.

Given the similarities between 
the structures of the differential cross section \rf{discross} 
and the result 
in Ref.\ \cite{klv17},
it is reasonable to expect that the strongest constraints
arise from low-$x$ data.
This suggests the best sensitivities are likely to involve 
the coefficients 
$a_f^{(5)}{}^{qqq}$, 
$a_f^{(5)}{}^{qqk}$, 
$a_f^{(5)}{}^{kqq}$, 
and $a_f^{(5)}{}^{kkq}$.
In principle,
data from the H1 and ZEUS collaborations at HERA
\cite{h1zeus}
and perhaps from a future electron-ion collider 
\cite{ls18}
could be analyzed to obtain constraints
on the various coefficients $a^{(5)}{}^{\mu\al\be}$.
Given that the HERA energies lie in the range 1-100 GeV,
the estimated constraints obtained 
in Ref.\ \cite{klv17}
via theoretical simulation of the HERA experiments
suggest it is feasible to achieve competitive sensitivities 
of order $10^{-7}$-$10^{-4}$ GeV$^{-1}$
to the coefficients $a^{(5)}{}^{\mu\al\be}$.
Verifying this by direct simulation of the experimental effects 
predicted by the differential cross section \rf{discross} 
would be worthwhile but lies beyond our present scope.

Since the laboratory frame for the DIS process is noninertial,
experimental results must be expressed in an inertial frame.
Most coefficients in the noninertial laboratory frame 
differ from those in the canonical inertial Sun-centered frame
\cite{sunframe}
by the time-dependent rotation \rf{rotz},
due to the Earth's rotation about its axis.
Nonetheless,
the differential cross section contains a time-independent part. 
The discussion in Sec.\ \ref{Lorentz-invariant limit}
shows that the coefficients $a^{(5)}{}^{\mu\al\be}$
cannot contain a Lorentz-invariant piece,
but the analysis in Sec.\ \ref{isotropic}
implies they do incorporate the three isotropic combinations
$a^{(5)TTT}, a^{(5)Tjj}, a^{(5)jjT}$
in each of the quark and proton sectors.
These isotropic components can generate 
a time-independent contribution to the differential cross section.
Given that the Earth rotates about the $Z$ axis,
we can expect this contribution to be expressed 
in the Sun-centered frame
in terms of the coefficient components
$a^{(5)TZZ}$, $a^{(5)TTZ}$, $a^{(5)TTT}$, 
$a^{(5)ZTT}$, $a^{(5)ZZT}$, $a^{(5)ZZZ}$, 
$a^{(5)TXX}+a^{(5)TYY}$, $a^{(5)ZXX}+a^{(5)ZYY}$,
$a^{(5)XXT}+a^{(5)YYT}$, $a^{(5)XXZ}+a^{(5)YYZ}$.

The time-dependent part of the differential cross section 
involves contractions of the coefficients $a^{(5)}{}^{\mu\al\be}$
with three four-momenta.
It therefore contains harmonics of the sidereal time
up to third order in the Earth's sidereal frequency
$\om_\oplus\simeq 2\pi/(23{\rm ~h} ~56{\rm ~min})$
\cite{ak98}
and can be expanded in the form 
\bea
&&
\hskip-10pt
\si(T_\oplus,x,Q^2)=\si_{\textrm SM}(x,Q^2)
\Big(1+a^{(5)\la\mu\nu}_{f\pr}\al^{f\pr}_{\la\mu\nu}
\nn\\
&&
\hskip10pt
+a^{(5)\la\mu\nu}_{f\pr}\be^{f\pr}_{\la\mu\nu}\cos\om_\oplus T_\oplus
+a^{(5)\la\mu\nu}_{f\pr}\ga^{f\pr}_{\la\mu\nu} \sin\om_\oplus T_\oplus
\nn\\
&&
\hskip10pt
+a^{(5)\la\mu\nu}_{f\pr}\de^{f\pr}_{\la\mu\nu}\cos2\om_\oplus T_\oplus
+a^{(5)\la\mu\nu}_{f\pr}\ep^{f\pr}_{\la\mu\nu} \sin2\om_\oplus T_\oplus
\nn\\
&&
\hskip10pt
+a^{(5)\la\mu\nu}_{f\pr}\va^{f\pr}_{\la\mu\nu}\cos3\om_\oplus T_\oplus
+a^{(5)\la\mu\nu}_{f\pr}\ze^{f\pr}_{\la\mu\nu} \sin3\om_\oplus T_\oplus\Big),
\nn\\
\eea
where $\al^{f\pr}_{\la\mu\nu}$, $\be^{f\pr}_{\la\mu\nu}$,
$\ga^{f\pr}_{\la\mu\nu}$, $\de^{f\pr}_{\la\mu\nu}$,
$\ep^{f\pr}_{\la\mu\nu}$, $\va^{f\pr}_{\la\mu\nu}$,
$\ze^{f\pr}_{\la\mu\nu}$ are functions of $x$ and $Q^2$,
and summation over the flavors $f\pr=f$ and $p$ is assumed.
The explicit forms of these functions 
can be obtained from Eq.\ \rf{discross}.
Note that the appearance of third harmonics
is a qualitatively new feature relative to the results 
in Ref.\ \cite{klv17}.

Finally,
we remark that the coefficients $a^{(5)}{}^{\mu\al\be}$
govern CPT-odd operators,
so their contribution to the differential cross section change sign
if the proton is replaced with an antiproton.
This can also be verified by an explicit calculation 
of the antiproton version of Eq.\ \rf{wmn}.
Since the coefficients belong to the quark and proton sectors,
they have no effects on the lepton or photon.
Also,
neglecting masses implies that the projectors
for electrons and positrons are equal, 
$\sum_s u^{(s)}(k)\overline{u}^{(s)}(k)
=\sum_s v^{(s)}(k)\overline{v}^{(s)}(k)$,
so the electron tensor $L_{\mu\nu}|_{e^-}$ 
and positron tensor $L_{\mu\nu}|_{e^+}$ are identical.
As a result,
replacing the electron with a positron
has no effect in our calculation.
The four possibilities for the cross section
are therefore related by
\bea
\frac{d\si}{dxdyd\ph}\bigg|_{e^- p,a}
&=&
\frac{d\si}{dxdyd\ph}\bigg|_{e^+ p,a}
\nn\\
&=&
\frac{d\si}{dxdyd\ph}\bigg|_{e^- \overline{p},-a}
=\frac{d\si}{dxdyd\ph}\bigg|_{e^+ \overline{p},-a}.
\qquad
\eea
Any differences between these cross sections
could in principle be used to isolate
effects from the coefficients $a^{(5)}{}^{\mu\al\be}$
and hence as a direct test of CPT symmetry.

\section{Summary}

This work investigates Lorentz- and CPT-violating operators 
in gauge field theories.
We construct gauge-invariant terms 
of arbitrary mass dimension $d$ in the Lagrange density 
describing fermions interacting with nonabelian gauge fields.
The construction is based on a technical result,
demonstrated in Sec.\ \ref{Gauge-covariant operators},
that any gauge-covariant combination 
of covariant derivatives and gauge-field strengths
can be written in the standard form \rf{form}.

The form of a generic gauge-invariant term in the Lagrange density
is discussed in Sec.\ \ref{Gauge Field Theory}.
Explicit expressions for all terms in the spinor sector with $d\leq 6$
are given in Table \ref{tab:spinor},
while all terms in the pure-gauge sector with $d\leq 8$
are displayed in Table \ref{tab:pure-gauge}.
We then discuss several interesting limiting cases
of the general formalism.
One is Lorentz-violating QED,
for which all operators with $d\leq 6$ are collected in Table \ref{tab:QED}.
Another is the Lorentz-violating theory of QCD and QED
with multiple flavors of quarks,
which has terms with $d\leq 6$ compiled in Table \ref{tab:QCD}.
A third limit of interest is the Lorentz-invariant case,
where the relevant operators are presented in Table \ref{tab:LI}.
We also consider the situation where Lorentz violation
is isotropic in a specified frame,
restricting attention to the operators listed  
in Table \ref{tab:isotropic}.

To illustrate the application of the results,
we study two experimental scenarios in Sec.\ \ref{Experiments}.
First,
corrections are calculated
to the cross section for light-by-light scattering
arising from nonminimal operators appearing at tree level 
in Lorentz-violating QED.
The results are combined with experimental data obtained at the LHC 
to place first constraints on 126 nonlinear operators with $d=8$,
collected in Tables \ref{tab:lls_iso} and \ref{tab:lls_one}.
Second,
we determine the modifications to the cross section for DIS 
arising from certain nonminimal Lorentz- and CPT-violating operators
in the theory of QCD and QED coupled to quarks.
The expression \rf{discross} for the differential cross section
suggests that an analysis of existing data
has the potential to place first constraints
on the corresponding nonminimal quark-sector coefficients.

The framework developed here encompasses a large variety of physical effects,
making them accessible to quantitative theoretical analysis.
It is evident that many avenues 
for phenomenological and experimental investigation
of realistic gauge field theories remain open for future investigation,
with a definite potential for the discovery of novel physical effects.

\section{Acknowledgments}

This work was supported in part
by the U.S.\ Department of Energy under grant {DE}-SC0010120
and by the Indiana University Center for Spacetime Symmetries.

\end{document}